\documentclass[ALICE,manyauthors]{cernphprep}
\usepackage[comma,square,numbers,sort&compress]{natbib}

\usepackage{hyperref}
\usepackage{xcolor}
\usepackage{lineno}
\newcommand{\XeXe}         {\mbox{Xe--Xe}\xspace}
\newcommand{\PbPb}         {\mbox{Pb--Pb}\xspace}

\newcommand{\nineH}        {$\sqrt{s}~=~0.9$~Te\kern-.1emV\xspace}
\newcommand{\seven}        {$\sqrt{s}~=~7$~Te\kern-.1emV\xspace}
\newcommand{\twoH}         {$\sqrt{s}~=~0.2$~Te\kern-.1emV\xspace}
\newcommand{\twosevensix}  {$\sqrt{s}~=~2.76$~Te\kern-.1emV\xspace}
\newcommand{\five}         {$\sqrt{s}~=~5.02$~Te\kern-.1emV\xspace}
\newcommand{\twosevensixnn}{$\sqrt{s_{\mathrm{NN}}}~=~2.76$~Te\kern-.1emV\xspace}
\newcommand{\fivenn}       {$\sqrt{s_{\mathrm{NN}}}~=~5.02$~Te\kern-.1emV\xspace}

\newcommand{\GeVc}         {Ge\kern-.1emV/$c$\xspace}
\newcommand{\MeVc}         {Me\kern-.1emV/$c$\xspace}
\newcommand{\TeV}          {Te\kern-.1emV\xspace}
\newcommand{\GeV}          {Ge\kern-.1emV\xspace}
\newcommand{\MeV}          {Me\kern-.1emV\xspace}
\newcommand{\GeVmass}      {Ge\kern-.2emV/$c^2$\xspace}
\newcommand{\MeVmass}      {Me\kern-.2emV/$c^2$\xspace}

\let\oldTeV\TeV 
\let\oldGeV\GeV
\renewcommand\TeV{\ensuremath\mathrm{\oldTeV}}
\renewcommand\GeV{\ensuremath\mathrm{\oldGeV}}
\newcommand\pT{\ensuremath p_{\scriptscriptstyle\mathrm{T}}}
\newcommand\sNN{\ensuremath\sqrt{s_{\scriptscriptstyle\mathrm{NN}}}}
\newcommand\cm{\ensuremath{\mathrm{cm}}}
\newcommand\vn[1]{\ensuremath v_{#1}}
\newcommand\vnm[3][]{\ensuremath v#1_{#2}\{#3\}}
\newcommand\rnn[1]{\ensuremath r_{#1|#1}}
\newcommand\mean[1]{\ensuremath\left\langle #1\right\rangle}

\newcommand\FigRef[1]{Fig.~\ref{#1}}
\newcommand\FigureRef[1]{Figure~\ref{#1}}

\newcommand\AMPT{AMPT}
\newcommand\Hydro{CLVisc}
\newcommand\OIO{POI\xspace}
\newcommand\RO{RFP\xspace}
\mathchardef\mhyphen="2D 
\newcommand\etaGap{\eta\mathrm{\mhyphen{}gap}}
\newcommand\etagap{\ifmmode\etaGap\else$\etaGap$\fi\xspace}

\newlength{\figwidth}
\def\paragraph#1{}
\def\subsection#1{}
\def\FIXME#1{}

\usepackage[T1]{fontenc}
\usepackage{orcidlink}

\begin{document}%

\begin{titlepage}
  \PHyear{2023}%
  \PHnumber{134}%
  \PHdate{12 July}%
  \title{Pseudorapidity
    dependence of anisotropic flow and its decorrelations using
    long-range multiparticle correlations in \PbPb{} and \XeXe{}
    collisions}%
  \ShortTitle{Pseudorapidity dependence of anisotropic
    flow} 
  \Collaboration{ALICE Collaboration\thanks{See
      Appendix~\ref{app:collab} for the list of collaboration
      members}}%
  \ShortAuthor{ALICE Collaboration}

\begin{abstract}

  The pseudorapidity dependence of elliptic ($\vn{2}$), triangular
  ($\vn{3}$), and quadrangular ($\vn{4}$) flow coefficients of charged
  particles measured in \PbPb{} collisions at a centre-of-mass energy
  per nucleon pair of $\sNN=5.02\,\TeV$ and in \XeXe{} collisions at
  $\sNN=5.44\,\TeV$ with ALICE at the LHC are presented. The
  measurements are performed in the pseudorapidity range
  $-3.5 < \eta < 5$ for various centrality intervals using two- and
  multi-particle cumulants with the subevent method. The flow
  probability density function (p.d.f.) is studied with the ratio of
  flow coefficient $v_2$ calculated with four- and two-particle
  cumulant, and suggests that the variance of flow p.d.f. is
  independent of pseudorapidity. The decorrelation of the flow vector
  in the longitudinal direction is probed using two-particle
  correlations. The results measured with respect to different
  reference regions in pseudorapidity exhibit differences, argued to
  be a result of saturating decorrelation effect above a certain
  pseudorapidity separation, in contrast to previous publications
  which assign this observation to non-flow effects.  The results are
  compared to $3+1$ dimensional hydrodynamic and the \AMPT{} transport
  model calculations.  Neither of the models is able to simultaneously
  describe the pseudorapidity dependence of measurements of
  anisotropic flow and its fluctuations. The results presented in this
  work highlight shortcomings in our current understanding of initial
  conditions and subsequent system expansion in the longitudinal
  direction. Therefore, they provide input for its improvement.

\end{abstract}
\end{titlepage}

\setcounter{page}{2} 

\section{Introduction}

There is significant evidence for the production of strongly coupled
plasma of quarks and gluons (QGP) in ultra relativistic heavy-ion
collisions, as measured by RHIC and LHC
experiments~\cite{BRAHMS:2004adc,PHOBOS:2004zne,STAR:2005gfr,PHENIX:2004vcz,ALICE:2022wpn}.
Several probes are used to determine the properties of this medium,
with measurements of anisotropic flow being among the most powerful
ones~\cite{Ollitrault:1992bk}. The nuclear overlap region of two
colliding nuclei forms an initial spatial anisotropy, which is
transformed, during the expansion of the subsequently created medium,
into an anisotropic azimuthal particle distribution. This anisotropy
is quantified based on the Fourier transform of the azimuthal particle
distribution~\cite{Voloshin:1994mz}
\begin{linenomath}
  \begin{align*}
    \frac{\mathrm{d}N}{\mathrm{d}\varphi}
    &\propto f(\varphi)
      = \frac{1}{2 \pi} \left[1 + 2 \sum_{n=1}^\infty
      \vn{n} \cos \left(n \left[\varphi - \Psi_n\right]\right)\right]
      \quad,
  \end{align*}
\end{linenomath}
where $\varphi$ is the azimuthal angle of the emitted particles,
$\Psi_n$ is the $n^{\text{th}}$ order flow symmetry plane and
$\vn{n} = \mean{ \cos \left(n \left[\varphi- \Psi_n\right] \right)}$
the $n^{\text{th}}$ anisotropic flow coefficient. Here,
$\mean{\ldots}$ denotes an average over all particles in a single
event. Together the $\vn{n}$ and $\Psi_n$ define the $n^{\text{th}}$
order (complex) anisotropic flow $V_n \equiv v_n e^{in\Psi_n}$, with
$v_n = |V_n|$ representing the magnitude of $V_n$ and $\Psi_n$ its
angle.

Anisotropic flow characterises the degree of collective motion of
produced particles relative to the symmetry plane vector of a
heavy-ion collision. It arises as a direct response to the initial
geometry of the overlapping region of colliding nuclei, expressed in
terms of eccentricity $\epsilon_n$ for
$n \leq 4$~\cite{Gardim:2012im,Heinz:2013bua}.  The most pronounced
component is the elliptic flow, $V_2$, related to the collision
ellipticity that reflects the almond shape of the nucleus overlap,
while higher order harmonics appear as a result of event-by-event
fluctuations of the initial transverse density profiles. Anisotropic
flow measurements have been studied in great detail both
experimentally and theoretically, thereby allowing the determination
of crucial information on the initial conditions and the transport
properties of the QGP~\cite{Heinz:2013th, Luzum:2013yya,
  Shuryak:2014zxa, Song:2017wtw}, such as the shear viscosity over
entropy density ratio, $\eta/s$, which was found to be near to the
universal lower bound of $1/4\pi$~\cite{Kovtun:2004de}. In these
studies, anisotropic flow was usually assumed to be driven by a boost
invariant initial spatial anisotropy, and experimental measurements
were interpreted as anisotropy with respect to an event-averaged
symmetry plane.

This assumption has been challenged by several measurements of the
pseudorapidity ($\eta$) dependence of anisotropic flow that revealed
longitudinal fluctuations of the flow vectors
$V_n$~\cite{Acharya:2017ino,CMS:2015xmx,ATLAS:2017rij,ATLAS:2020sgl}.
This can be interpreted as decorrelation of the flow magnitudes and/or
symmetry plane angles between two different $\eta$
windows. Measurements exploiting multiparticle correlations suggest
that fluctuations in both the flow magnitude and the symmetry plane
contribute equally to the flow vector
decorrelation~\cite{ATLAS:2017rij}. It was argued in
Ref.~\cite{Jia:2014ysa} that these effects are connected to the
fluctuating initial state, where the transverse shape of the initially
produced system fluctuates not only on an event-by-event basis, but
also within an event, and it depends on $\eta$. Indeed, many
theoretical studies based on
hydrodynamic~\cite{Bozek:2015bna,Pang:2015zrq,Bozek:2017qir} and
transport models~\cite{Jia:2014ysa,Xiao:2012uw} showed that the
decorrelations are connected to the longitudinal fluctuations in the
initial state, with a possible additional contribution from early time
hydrodynamic fluctuations~\cite{Sakai:2020pjw}, but only weak
dependence on the $\eta/s$ of the
QGP~\cite{Pang:2015zrq,Sakai:2020pjw}.  Measurements of anisotropic
flow and its fluctuations as a function of pseudorapidity, therefore,
represent an important ingredient to constrain the three-dimensional
initial conditions and the QGP expansion in longitudinal
direction~\cite{Denicol:2015nhu,Molnar:2014zha,Shen:2017bsr}.

It was found that comparison between measurements from \PbPb{} and
\XeXe{} collisions offers a unique possibility to test the hydrodynamic
framework under variations of the nuclear mass number and geometry of
the
collisions~\cite{Eskola:2017bup,Giacalone:2017dud,ALICE:2018lao}. Recent
results on longitudinal flow fluctuations in both \PbPb{} and \XeXe{}
collisions showed that the hydrodynamic models that successfully
describe the transverse dynamics of the medium evolution, do not
reproduce the longitudinal structure of the initial
state~\cite{ATLAS:2020sgl}. Thus, studying results from collision
systems of different sizes can bring additional insight into our
understanding of the properties of the QGP.

This letter presents measurements of the pseudorapidity dependence of
anisotropic flow coefficients $\vn{2}$, $\vn{3}$ and $\vn{4}$ in
\PbPb{} collisions at collision energy $\sNN = 5.02\,\TeV$ and \XeXe{}
collisions at $\sNN = 5.44\,\TeV$ within a wide pseudorapidity range,
$-3.5 < \eta < 5.0$, with the ALICE detector. The measurements are
based on two- and four-particle cumulants, $\vnm{n}{2}$ and
$\vnm{n}{4}$,
respectively~\cite{Borghini:2000sa,Bilandzic:2013kga}. To suppress
contamination from short-range correlations, denoted as non-flow,
particles are measured in different subevents widely separated in
phase space by imposing a pseudorapidity gap, $|\Delta\eta|$, between
them.  The large detector acceptance, together with the
state-of-the-art measurement techniques, enables us to perform these
studies in a wider pseudorapidity range and with a larger
pseudorapidity separation of $|\Delta\eta| > 3.8$ compared to previous
measurements done at the LHC~\cite{CMS:2015xmx, ATLAS:2017rij,
  CMS:2017xnj}. The results presented in this letter improve previous
ALICE measurements~\cite{ALICE:2016tlx} by significantly reducing the
systematic uncertainties dominated by those stemming from corrections
for secondary particles and non-flow contamination.  In addition, the
longitudinal flow vector fluctuations are investigated using the ratio
of two-particle correlators calculated in subevents at different
pseudorapidities. Both centrality and pseudorapidity dependences are
discussed. The results are compared with calculations from a $3+1$
dimensional \Hydro{} hydrodynamic model~\cite{Wu:2018cpc} and the
\AMPT{} transport model~\cite{Lin:2004en}.

\section{Observable definitions}
\label{sec:observables}

Anisotropic flow and its fluctuations are measured with two- and
four-particle cumulants, and a decorrelation ratio, respectively, with
the use of $m$-particle azimuthal correlations.  The $m$-particle
azimuthal correlations are calculated with the generic
framework~\cite{Bilandzic:2013kga}, which is an effective way to
obtain correlation of any order corrected for detector effects.

Differential observables, such as those presented in this work, are
measured by correlating the so-called particles of interest, \OIO{},
in the desired narrow pseudorapidity interval with respect to
reference particles, \RO{}, measured in a wide pseudorapidity
range. If the $\eta$ intervals of \OIO{} and \RO{} are close to each
other, such correlations are affected by non-flow contamination, that
mainly arise from correlations among the jet constituents. These are
suppressed by imposing a pseudorapidity gap, $|\Delta\eta|$, between
regions called subevents. In this analysis, correlations are
calculated in specific regions in $\eta$, which can naturally be
considered as subevents. The choice of these subevents allows us to
perform longitudinal flow measurements while suppressing the
contribution from non-flow.

\begin{figure}[htbp]
  \centering
  \includegraphics[width=0.85\figwidth]{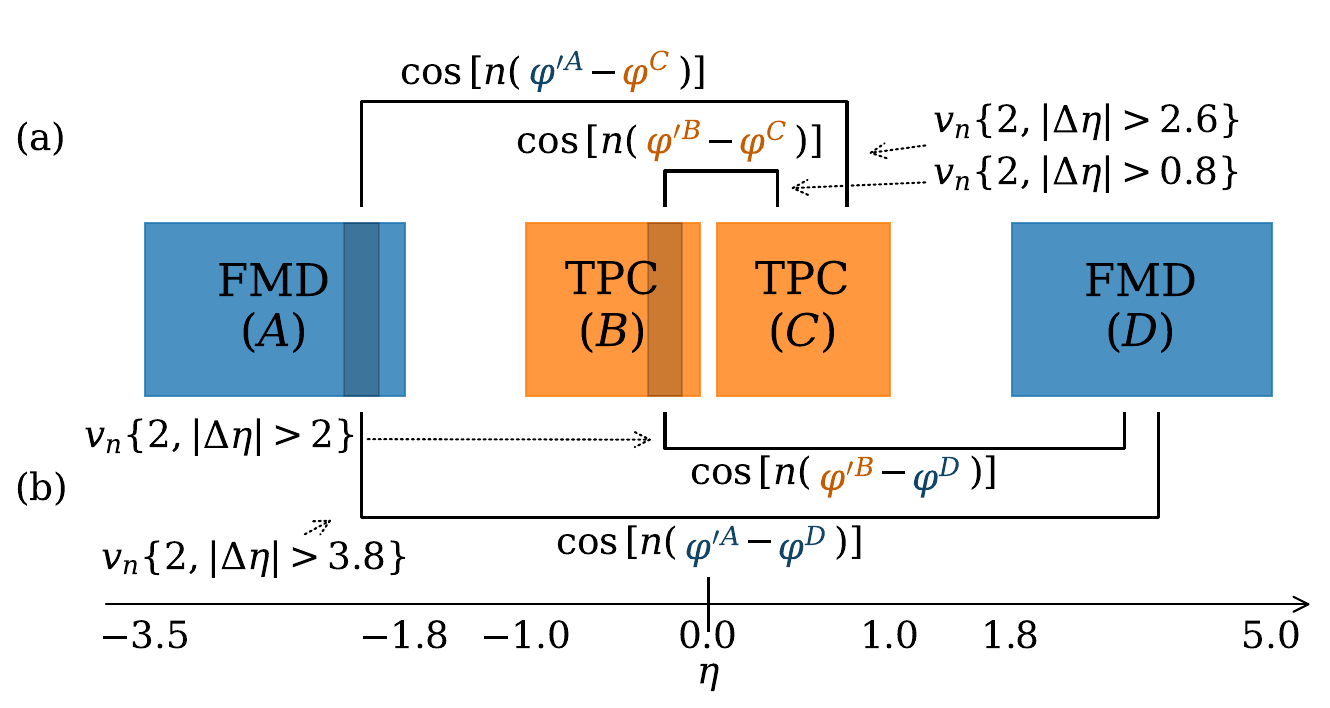}
  \caption[Illustration of correlation methods 1.]{Illustration of
    correlator methods showing calculation of $\vnm{n}{2}$ using
    either a small (a) or large (b) separation in pseudorapidity.
    Darker bands indicate where the differential measurement is
    performed (i.e. particles of interest) while the other end of the
    connecting lines indicate reference particles.}
  \label{fig:cumulants:method1}
\end{figure}

\begin{figure}[htbp]
  \centering
  \includegraphics[width=0.85\figwidth]{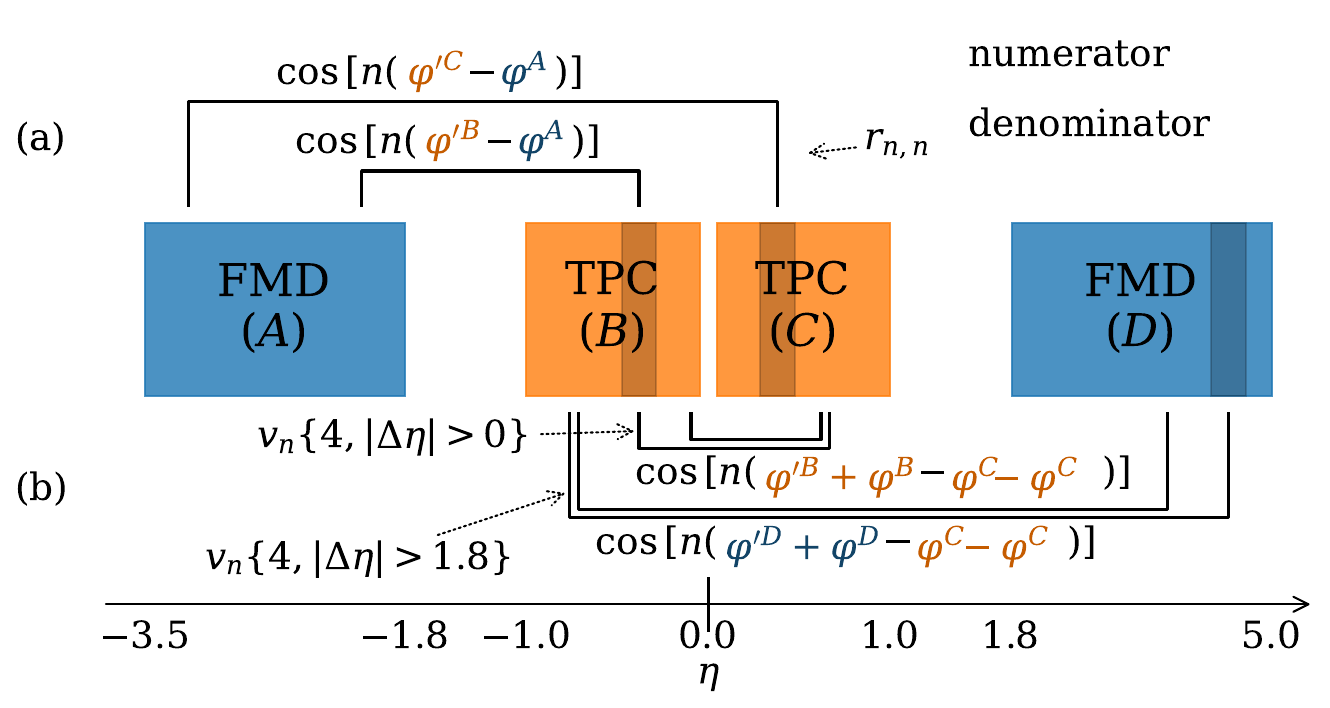}
  \caption[Illustration of correlation methods 2.]{Illustration of
    correlator methods showing calculation of the decorrelation effect
    (a) as well as $\vnm{n}{4}$ (b).  Darker bands indicate where the
    differential measurement is performed (i.e. particles of interest)
    while the other end of the connecting lines indicate reference
    particles.}
  \label{fig:cumulants:method2}
\end{figure}

The regions used for the measurements presented in this article are
schematically illustrated in Figs.~\ref{fig:cumulants:method1}
and~\ref{fig:cumulants:method2}.  Regions $A$ and $D$ correspond to
very forward pseudorapidities, while regions $B$ and $C$ are on either
side of the symmetry line $\eta=0$.

Within a traditional approach, flow coefficients differential in
pseudorapidity are obtained from two-particle cumulants as
\begin{linenomath}
  \begin{align}\label{eq:tpcref}
    \vnm[^{\prime X}]{n}{2}
    & = \frac{\langle \langle 2^{\prime}\rangle\rangle}{
      \sqrt{\langle \langle 2 \rangle\rangle}}
      = \frac{\mean{\vn{n}^{\prime X}\vn{n}^Y}}{
      \sqrt{\mean{\vn{n}^X\vn{n}^Y}}},
  \end{align}
\end{linenomath}
where the $v_n$ and $v_n^{\prime}$ are the reference and differential
flow, respectively, and $X$, $Y$ stand for different reference regions
for particle correlations in pseudorapidity. The single angular
brackets $\langle \cdot \rangle$ represent an average over events with
similar centrality, and the double angular brackets
$\langle\langle \cdot \rangle\rangle$ an average over particles within
an event, and over events. It is further assumed that the reference
flow is symmetric, $v_n^X = v_n^Y$, as warranted for symmetric
collision systems such as Pb--Pb and Xe--Xe, presented in this
work. The $\langle\langle m\rangle\rangle$ denotes the $m$-particle
correlation, in particular the $\langle\langle 2\rangle\rangle$
represents correlations between two \RO{}, and
$\langle\langle 2^{\prime}\rangle\rangle$ correlation between \RO{}
and \OIO{}. These correlations are defined as:
\begin{linenomath}
  \begin{align}\label{eq:twoptcl}
  \begin{split}
    \langle\langle 2\rangle\rangle
    & = \langle \langle {\rm cos}[{\rm n}(\varphi_1^{X}
    - \varphi_2^{Y})] \rangle\rangle,\\
    \langle\langle 2^{\prime}\rangle\rangle
    & = \langle \langle {\rm cos}[{\rm n}(\varphi_1^{\prime X}
    - \varphi_2^{Y})] \rangle\rangle,
      \end{split}
  \end{align}
\end{linenomath}
with $\varphi_k^{X,Y}$ representing the azimuthal angle of \RO{} from
reference regions $X$ or $Y$, and $\varphi_k^{\prime X}$ representing
the azimuthal angle of \OIO{} in a given narrow interval in $\eta$.

\paragraph{Cumulants with modest pseudorapidity separation}

The analysis is carried out with two choices of reference region, as
illustrated in \FigRef{fig:cumulants:method1}.  In the first case, the
reference region is chosen from midrapidity, i.e. either region $B$ or
$C$. In this configuration, the \RO{} are correlated with \OIO{} in
region $C$ or $D$, or $B$ or $A$, respectively. That is, \RO{} at
negative (positive) midrapidity are correlated with \OIO{} at positive
(negative) mid or forward rapidity. In such configurations where the
correlated particles are taken from neighbouring regions, it is
important to suppress non-flow by avoiding correlations near the edge
of the regions, i.e. when the difference between \OIO{} and \RO{} is
$|\Delta\eta| \approx 0$. Therefore, only a specific $\eta$ range in
the corresponding reference region is chosen, in particular
$0.8 < |\eta| < 1.0$. This choice effectively results in an \etagap{}
for midrapidity measurements of $|\Delta\eta| > 0.8$, while forward
measurements have a larger separation of $|\Delta\eta| > 2.6$.

The case where the reference region is positioned in $C$ is
illustrated in \FigRef{fig:cumulants:method1}(a).

\paragraph{Cumulants with large pseudorapidity separation}

In the second case, the reference region is chosen to be at forward
rapidity. In this configuration, there are again several options to
correlate the \RO{} with \OIO{}. The particles taken from reference
regions $A$ (or $D$), can be correlated with \OIO{} from $C$ or $D$
(or $A$ or $B$). In this way, the pseudorapidity separation between
\RO{} and \OIO{} is increased to $|\Delta\eta| > 2.0$ and
$|\Delta\eta| > 3.8$ for mid- and forward rapidity measurements,
respectively. Therefore, this configuration excludes more short-range
correlations arising from non-flow as compared to the first case. This
configuration, in particular the case where the reference region is
chosen to be $D$, is illustrated in \FigRef{fig:cumulants:method1}(b)
.

\paragraph{Multiparticle cumulants}

The advantage of measuring flow coefficients via $m$-particle
cumulants for $m>2$ is the suppression of lower order non-flow
correlations by definition~\cite{Borghini:2000sa}, including those
that stem from non-flow effects. To further suppress remaining
non-flow originating in multiparticle short-range correlations, the
subevent method was recently introduced to higher order cumulants,
too~\cite{Jia:2017hbm,Huo:2017nms}. The differential flow is
determined using the four-particle cumulant defined as
\begin{linenomath}
  \begin{align}\label{eq:4cum}
    \vnm[^{\prime X}]{n}{4}
    &=
      -\frac{\langle\langle 4^{\prime}\rangle\rangle
      - 2\cdot \langle\langle 2^{\prime} \rangle\rangle
      \langle\langle 2\rangle\rangle}{(- \langle\langle 4\rangle\rangle
      - 2\cdot \langle\langle 2\rangle\rangle^2)^{3/4}}
      = \frac{\langle v_n^{\prime X}v_n^{3}\rangle}{
      \langle v_n^{4}\rangle^{3/4}},
  \end{align}
\end{linenomath}
where the two-particle correlations are calculated in the same way as
in Eq.~\eqref{eq:twoptcl}. Four-particle correlations are obtained
as
\begin{linenomath}
  \begin{align}\label{eq:fourptcl}
  \begin{split}
   \langle\langle 4\rangle\rangle
    &= \langle\langle {\rm cos} [n(\varphi_1^X + \varphi_2^X - \varphi_3^Y - \varphi_4^Y)]\rangle\rangle, \\
    \langle\langle 4^{\prime}\rangle\rangle
    &= \langle\langle {\rm cos} [n(\varphi_1^{\prime X} + \varphi_2^X - \varphi_3^Y - \varphi_4^Y)]\rangle\rangle.
    \end{split}
  \end{align}
\end{linenomath}

The choice of the subevents for the $\vnm{n}{4}$ measurements is the
same as in the case of the two-particle cumulant measurement discussed
above and as illustrated in \FigRef{fig:cumulants:method2}(b).
Reference regions only at midrapidity, $B$ or $C$, are used for this
measurement. As this observable is less influenced by non-flow, it is
possible to exploit the whole $\eta$ range of the reference region to
minimise statistical uncertainties, yielding an \etagap{} of
$|\Delta\eta| > 0$ for midrapidity measurements, and
$|\Delta\eta| > 2.0$ for forward measurements.

\paragraph{decorrelation of flow vector}

In order to investigate the longitudinal fluctuations of the flow
vector, a decorrelation ratio $\rnn{n}$~\cite{CMS:2015xmx} is used.
As illustrated in \FigRef{fig:cumulants:method2}(a), it is formed as
the ratio of the opposite-side two-particle correlation between the
reference region and the region of interest (i.e. \RO{} from region
$A$ ($D$) correlated with \OIO{} in narrow $\eta$ intervals from
region $C$ ($B$)) to the same-side correlation (i.e. \RO{} from region
$A$ ($D$) correlated with \OIO{} in narrow $\eta$ intervals from
region $B$ ($C$)),
\begin{linenomath}
  \begin{align}\label{eq:rnn}
    \rnn{n}
    &= \frac{\langle{\rm cos}[n(\varphi_1^{\prime -X}
      - \varphi_2^Y)]\rangle}{\langle{\rm cos}[n(\varphi_1^{\prime X}
      - \varphi_2^Y)]\rangle}
    = \frac{\mean{\vn{n}^{\prime -X}\vn{n}^{Y}\cos [n
      \left(\Psi_n^{\prime -X}-\Psi_n^{Y}\right)]}}{ 
      \mean{\vn{n}^{\prime X}\vn{n}^{Y}\cos [n
      \left(\Psi_n^{\prime X}-\Psi_n^{Y}\right)]}}\quad,
  \end{align}
\end{linenomath}
where the $\Psi_n^{Y}$ is the average symmetry plane of a reference
region $Y$, and $\Psi_n^{\prime X}$ is the symmetry plane of the given
narrow interval in $\eta$. Flow vector fluctuations arise from two
sources. First, the decorrelation of the symmetry plane, which would
manifest in a non-vanishing cosine term, as the symmetry planes
$\Psi_n$ at different pseudorapidities would not be equal to each
other, or twisted, $\Psi_n^{X} \neq \Psi_n^{Y}$. Second, the
decorrelation of the flow magnitude would lead to the product of flow
coefficients not being factorised due to additional fluctuation terms
dependent on $\eta$:
$\langle v_n^Xv_n^Y\rangle \neq \sqrt{\langle
  {v_n^X}^2\rangle}\sqrt{\langle {v_n^Y}^2\rangle}$. To measure the
absolute signal of flow vector fluctuation, the correlations between
particles from different $\eta$ intervals in the numerator of
$\rnn{n}$ would be ideally divided with correlations performed between
particles from the same $\eta$ window in the denominator. Such
configuration would, however, introduce significant non-flow
background, as the main characteristic of these correlations is the
proximity of particles in $\eta$. Therefore, the $\rnn{n}$ observable
defined in Eq.~\eqref{eq:rnn} is used, which quantifies the relative
flow fluctuations between $\eta$ and $-\eta$. It ensures that \OIO{}
have the same absolute pseudorapidity ($\eta = \eta^C=-\eta^B$). If
neither $v_n$ nor $\Psi_n$ fluctuate along the longitudinal direction,
then $\rnn{n}$ is expected to converge to unity. If either of these
effects, or both, are present, then the numerator of $\rnn{n}$ is
smaller than the denominator, and hence $\rnn{n}$ will deviate from
unity, with deviations growing stronger with increasing $\eta$
(i.e. larger relative pseudorapidity difference). However, remaining
short-range non-flow effects may also give rise to a fake
decorrelation signal, as they would increase the value of the
denominator.

\section{Experimental Setup}
\label{sec:experiment}

\paragraph{Overview}

A detailed description of the ALICE apparatus can be found in
Ref.~\cite{ALICE:2008ngc,ALICE:2014sbx}. The relevant detectors for
the presented results are the Inner Tracking System
(ITS)~\cite{ALICE:2010tia}, a silicon detector consisting of 6
cylindrical layers close to the collision point; the Time Projection
Chamber (TPC)~\cite{Alme:2010ke}, which is the main tracking detector
in ALICE; and the Forward Multiplicity Detector
(FMD)~\cite{ALICE:2004ftm}, a silicon strip detector which measures
the multiplicity of charged particles at forward rapidities.  Finally,
the V0 scintillator arrays are used for online event selection and
offline centrality determination. These two arrays are placed at very
forward rapidities and provide high-resolution timing and approximate
sum multiplicity measurements offline~\cite{ALICE:2013axi}.

\paragraph{Midrapidity}

Charged-particle trajectories measured by both ITS and TPC combined
are limited to $|\eta|<0.9$.  However, tracks measured with the TPC
only, that is, without matching tracks to other hit-producing devices,
can be extended to $|\eta|<1.5$, with a slightly reduced transverse
momentum resolution.  Particle trajectories are reconstructed in
$0.2\,<\pT<5\,$~GeV/$c$, with the requirement of at least one space
point in the two inner most layers of the ITS, at least 70 (out of
159) space points in the TPC, and a largest transverse and
longitudinal distance (DCA) to the primary vertex of $0.0182\,\cm$ and
$2\,\cm$, respectively.  The upper cut-off on transverse momentum is
imposed to limit the contribution from high-$\pT$ particles mostly
from jets and to ensure relatively uniform tracking efficiency.  The
limit on pseudorapidity is imposed to ensure full TPC coverage.
Tracking efficiencies and detector acceptance effects are corrected
for using per-particle weights (for details see
Ref.~\cite{Bilandzic:2013kga}).  The flow measurements in the central
region of pseudorapidity are extrapolated to $\pT{} = 0$, based on
simulations with \AMPT{} model calculations as input, to make the
measurements comparable to those at forward pseudorapidity. The
correction was obtained as the ratio of $v_n$ with no $p_{\rm T}$ cut
to $v_n$ with the cut used for the analysis. Only primary particles
were used to remove any detector effects in the calculations.

\paragraph{Forward-rapidity}

The FMD covers $-3.5<\eta<-1.8$ and $1.8<\eta<5$ with high resolution
in pseudorapidity and 20 segments in azimuth. Due to various technical
and engineering considerations, the direct line-of-sight from the FMD
sub-detectors to the collision point is obscured by relatively large
amounts of material which require careful study of secondary particle
production and associated corrections~\cite{ALICE:2015bpk}.  The FMD
does not provide any tracking capabilities on its own and can
therefore not distinguish between primary
particles~\cite{ALICE-PUBLIC-2017-005} and particles produced in
decays or surrounding material.  Instead, the effect of secondary
particles on the observed azimuthal particle distribution was studied,
with emphasis on the relative deflection of secondary particles with
respect to the primary particle ($\delta\varphi$).  The Fourier
transform $\mathcal{F}$ of the resulting
$\mathrm{d}N/\mathrm{d}(\delta\varphi)$ distribution then corrects for
the deflection of secondary particles relative to their primary
origin, so that the differential flow measurement becomes
$\mean{\mean{v'_nv_n}}^{\mathrm{primary}}=\mathcal{F}^{-1}\mean{\mean{v'_nv_n}}^{\mathrm{inclusive}}$,
where \emph{inclusive} means the measured correlation of both primary
\emph{and} secondary particles. The transformation $\mathcal{F}$ (a
factor in Fourier space) depends on the material traversed by the
particles and, as such, is dependent on the pseudorapidity of the
particles and on the position of the interaction point in the beam
direction $z$. The first order effect of the material is to amplify
the primary signal proportionally, and no significant dependence of
$\mathcal{F}$ on collision centrality is found.  The resulting
correction for the effect of secondary particles ranges from 1.1 to
1.6, depending on the pseudorapidity, primary vertex $z$ coordinate,
and the order of the flow harmonic.  Detector acceptances are
corrected for using per-azimuth-segment weights.

\paragraph{Data samples}

The results presented in this letter are based on data acquired during
LHC Run~2 in 2015 for \PbPb{} collisions at $\sNN = 5.02\,\TeV$ and in
2017 for \XeXe{} at $\sNN = 5.44\,\TeV$. The event selection involves
a centrality estimate based on the amplitudes of the signals in both
arrays of the V0 detectors~\cite{ALICE:2013hur}, and a constraint on
events within 10 cm from the primary vertex. Selection criteria on the
correlation between the forward detectors V0 and FMD as well as a veto
on multi-vertex events are applied to select beam-crossings and reject
pileup as well as outlier events. In total, $10^7$ and $10^6$
collisions of \PbPb{} and \XeXe{}, respectively, were analysed.

\section{Systematic uncertainties}
\label{sec:uncertainties}

\paragraph{General}

The systematic uncertainties stemming from event selection,
multi-vertex veto, and outlier rejection are investigated by
tightening and relaxing the selection criteria used in this analysis,
and the effects are found to be negligible. Variations in the accepted
primary vertex $z$ coordinate influence the acceptance of the
detectors. This contribution accounts for at most $1\%$ uncertainty in
the most peripheral collisions. The systematic uncertainty associated
with estimating collision centrality is studied by defining centrality
intervals using the multiplicity distribution measured at
midrapidity~\cite{ALICE:2018tvk} rather than in the V0 amplitude.  The
uncertainty is found to be at most $1\%$ in midcentral collisions and
negligible in the most peripheral and central collisions.

\paragraph{Midrapidity}

The effects of the track selection at midrapidity are explored by
performing an independent analysis with varied values of the selection
criteria, and is summarised in the following. Increasing the number of
required TPC space points was found to be negligible. Variations in
the required DCA of tracks in both transverse and longitudinal
direction to provide different sensitivity to contamination from
secondary tracks result in a $1$ to $3\%$ systematic uncertainty,
larger for most central collisions and $\vnm{n}{4}$ results.
\emph{Hybrid tracks}, which combine information from three different
types of tracks in order to achieve uniform azimuthal acceptance and
the best transverse momentum resolution, are used for systematic
variation of the track reconstruction procedure.  A $1$ to $6\%$
systematic uncertainty is found, smallest for $v_2$, and largest for
the highest harmonic, $v_4$. The uncertainty arising from the
extrapolation of midrapidity $\vn{n}$ to $p_{\rm T}=0$ was estimated
by varying the maximum value of the $p_{\rm T}$ selection, and was
found to be $1\%$.

\paragraph{Forward rapidity}

Systematic uncertainties originating from secondary particle
production on a material in front of the FMD are investigated by
varying the material density by $\pm10\%$ using Monte Carlo
simulations.  The resulting systematic uncertainty of $2.5\%$ for
$\vn{2}$, $3\%$ for $\vn{3}$, and $3.5\%$ for $\vn{4}$, is found. This
represents a significant improvement with respect to the previous
results~\cite{ALICE:2016tlx}.  An effective correction for secondary
particle production, which compares generator level to post-simulated
detector response, to the Fourier space correction ($\mathcal{F}$),
gives a systematic uncertainty ranging from $1\%$ for $v_2$ to $4\%$
for $v_4$.  Finally, the systematic uncertainty from generating the
secondary correction on simulated rather than experimental data, via
detailed analysis of simulated particle trajectories, ranges between
0.5--3$\%$, with values dependent on the order of the harmonic
investigated.

As the effects from secondary particles arising from including the FMD
detector in the analysis has little or no dependence on centrality nor
collision systems, the same corrections for secondary particles in
\PbPb{} collisions are applied to the \XeXe{} data, taking into
account the smaller overall particle production in these
collisions~\cite{ALICE:2018cpu}.

The weights introduced to account for non-uniform acceptance and
efficiencies are in principle dependent on the granularity by which
these are determined.  The uncertainty related to varying the
granularity is investigated and found to be negligible at both mid-
and forward rapidity.

The systematic uncertainty of blind regions (`holes') in the FMD is
evaluated by comparing the generator level results to full detector
response simulations results with interpolation in these holes. This
exercise results in $2\%$ systematic uncertainty, applied only in the
pseudorapidity regions affected by the acceptance holes of the FMD.
Also, the positive and negative pseudorapidity results are compared,
as these are expected to be symmetric in symmetric collision systems
such as \PbPb{} and \XeXe{}.  An uncertainty of at most $2\%$ and
$4\%$ at mid- and forward rapidity, respectively, in \PbPb{}
collisions, and $2\%$ and $5\%$ in \XeXe{} collisions is assigned, due
to the asymmetry introduced into the analysis procedure by corrections
for efficiency, acceptance, and secondary particles.

It should be noted that the choice of using two- or multiparticle
correlations, with a rapidity gap between \OIO{} and \RO{} to suppress
correlations from non-collective behaviour, effectively removes the
2--10\% uncertainty that was applied to the previous ALICE
results~\cite{ALICE:2016tlx}.

The different sources listed above were assumed to be uncorrelated and
added in quadrature to determine the total systematic uncertainty on
the measurement of $v_n$. These contributions, however, cancel out in
the decorrelation ratio $\rnn{2}$, since uncertainties are shared
between the numerator and denominator. Therefore, only the statistical
uncertainties are reported on this quantity.

\section{Results}
\label{sec:results}

\subsection{Flow measurements in \PbPb{} collisions}

\paragraph{Modest pseudorapidity separation}

\begin{figure}[htbp]
  \centering
  \includegraphics[width=\figwidth]{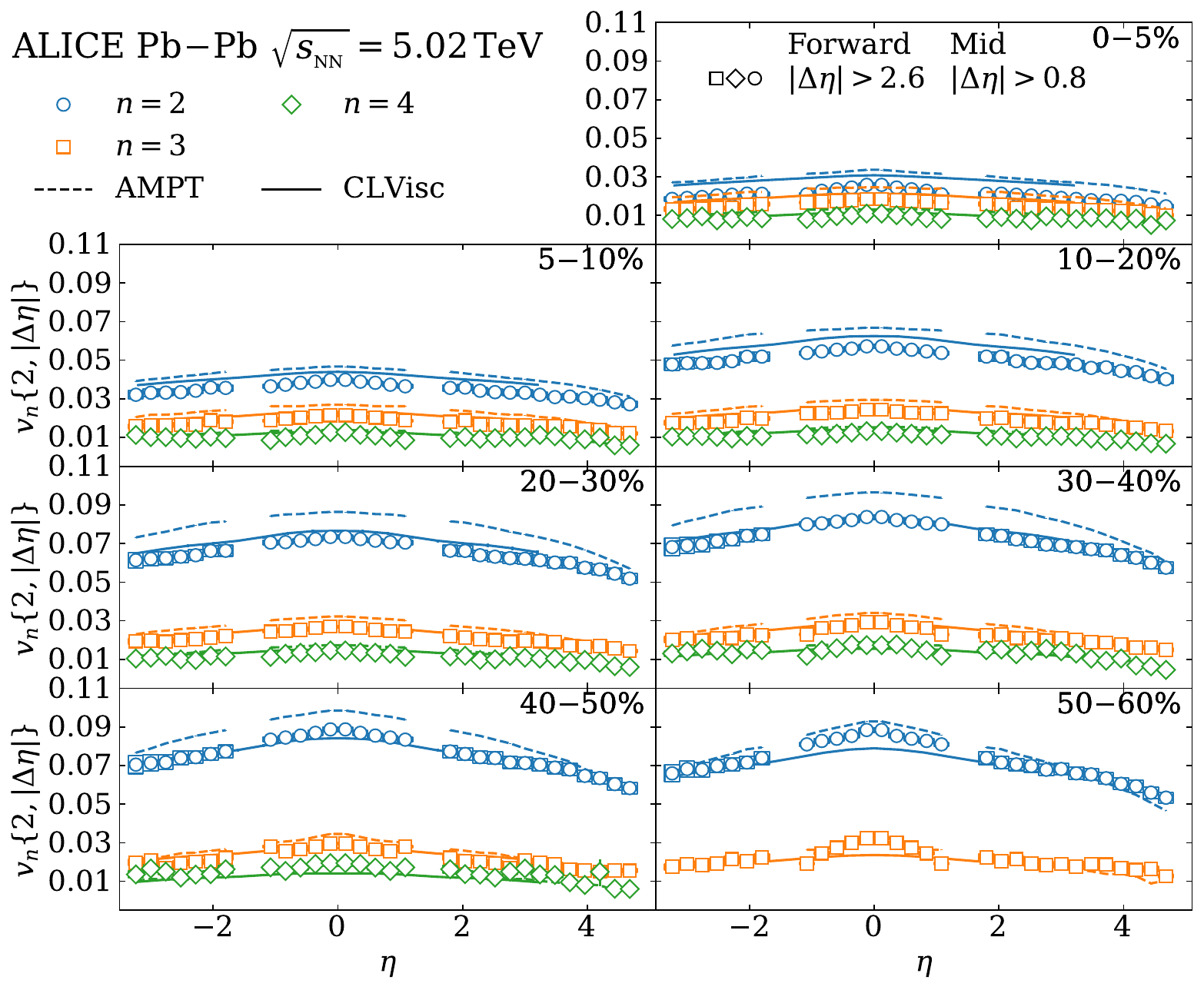}
  \caption[$\vnm{n}{2,|\Delta \eta| > 0.8}$ with TPC reference in
  \PbPb{} collisions.]{The differential flow measurements,
    $\vnm{n}{2}(\eta)$, measured with the 2-particle cumulant and
    chosing reference particles from the TPC in \PbPb{} collisions at
    $\sNN{} = 5.02\,\TeV{}$. The choice of the reference particles
    results in $|\Delta \eta | > 0.8$ and $2.6$ in the mid and forward
    pseudorapidity regions, respectively. At midrapidity, the results
    are extrapolated to $\pT=0$. \AMPT{} and \Hydro{} model
    calculations are compared with the results.}
  \label{fig:result:pbpb:gap08}
\end{figure}

\FigureRef{fig:result:pbpb:gap08} presents the measurements of
$\vnm{2}{2}$, $\vnm{3}{2}$ and $\vnm{4}{2}$ as a function of
pseudorapidity and centrality in \PbPb{} collisions at
$\sNN=5.02\,\TeV$. The $\vnm{4}{2}$ measurement at 50--60$\%$
collision centrality is not shown due to large statistical
uncertainties.  The reference region is at midrapidity, and the
\etagap{} between the \OIO{} and \RO{} at mid and forward rapidity is
$|\Delta \eta| > 0.8$ and $|\Delta \eta| > 2.6$, respectively
(corresponding to the subevent topology illustrated in
\FigRef{fig:cumulants:method1}(a)). The $\vnm{2}{2}$ measurements show
a stronger dependence on collision centrality, while $\vnm{3}{2}$ and
$\vnm{4}{2}$ reveal only a modest dependence. This is consistent with
pseudorapidity integrated measurements~\cite{Adam:2016izf} explained
by $v_2$ being driven by the average elliptic geometry, while higher
order harmonics originate predominantly from its fluctuations.

\paragraph{Large pseudorapidity separation, and four-particle correlations}

\begin{figure}[htbp]
  \centering
  \includegraphics[width=\figwidth]{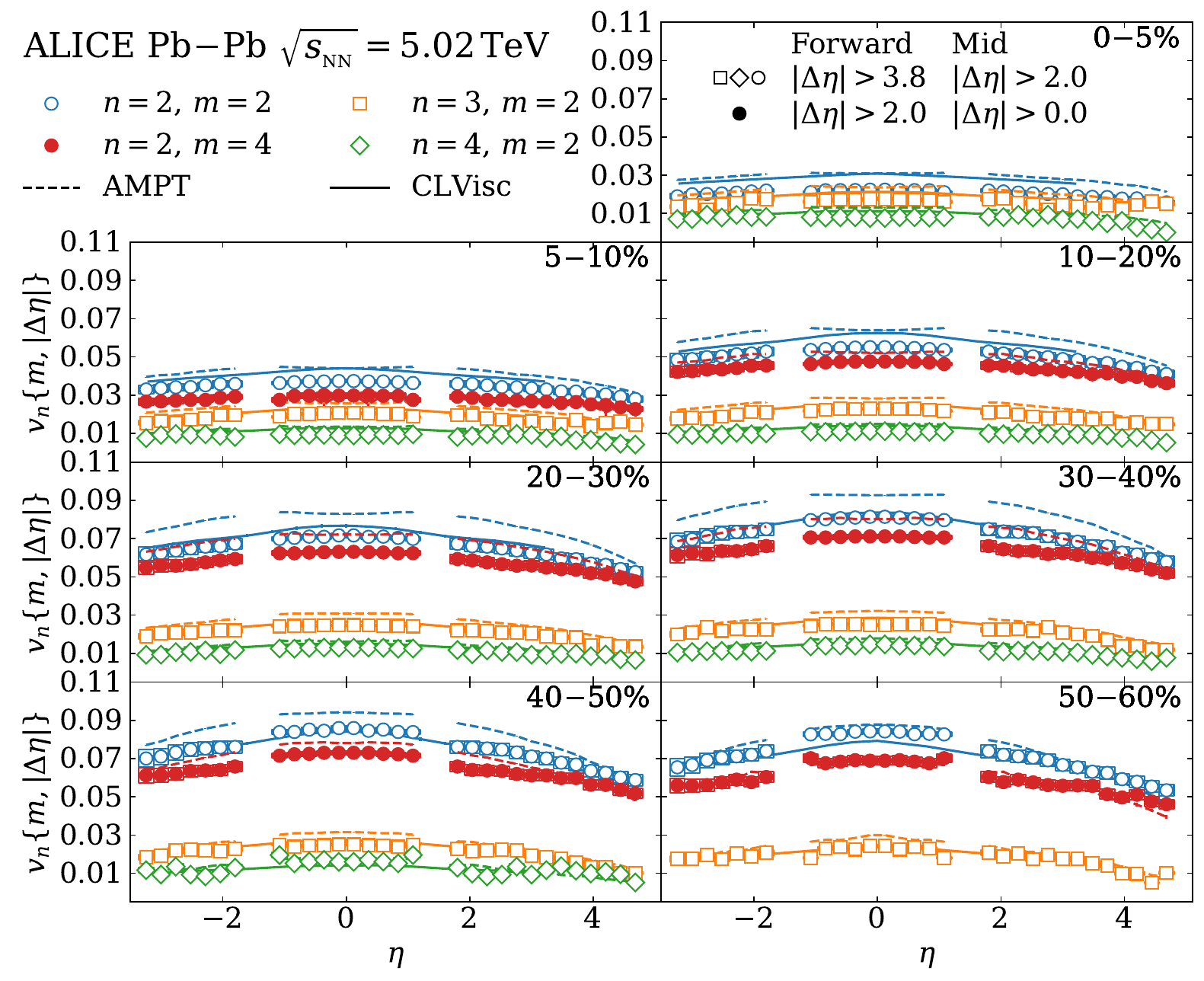}
  \caption{Two-particle cumulant with reference particles chosen from
    the FMD and 4-particle cumulant with reference particles chosen
    from the TPC in \PbPb{} collisions at $\sNN{} = 5.02\,\TeV{}$. The
    pseudorapidity separations are $|\Delta \eta | > 2.0$ and $3.8$
    for two-particle cumulants for mid and forward rapidities,
    respectively.  For the $\vnm{2}{4}$ the separations are
    $|\Delta \eta | > 0$ and $2.0$ for mid and forward rapidities,
    respectively.  At mid rapidity, the results are extrapolated
    to $\pT=0$. \AMPT{} and \Hydro{} model calculations are compared with
    the results.}
\label{fig:result:pbpb:gap20}
\end{figure}

\FigureRef{fig:result:pbpb:gap20} shows the pseudorapidity dependence
of flow coefficients obtained with large \etagap{}
($|\Delta\eta| > 2.0$ and $|\Delta\eta| > 3.8$ for mid and forward
rapidities, respectively) between \OIO{} and \RO{} choosing forward
rapidity for the reference region (corresponding to the subevent
topology illustrated in \FigRef{fig:cumulants:method1}(b)).  Also
shown are the results of the four-particle cumulant $\vnm{2}{4}$. A
similar trend of $v_n\{m\}$ as a function of centrality and
pseudorapidity can be seen as in \FigRef{fig:result:pbpb:gap08},
except at midrapidity, where the increased \etagap{} for $\vnm{n}{2}$,
or using higher order cumulant $\vnm{2}{4}$, leads to almost constant
dependence on pseudorapidty compared to the measurements with smaller
\etagap{} reported in \FigRef{fig:result:pbpb:gap08}.

\begin{figure}[htbp]
  \centering
  \includegraphics[width=\figwidth]{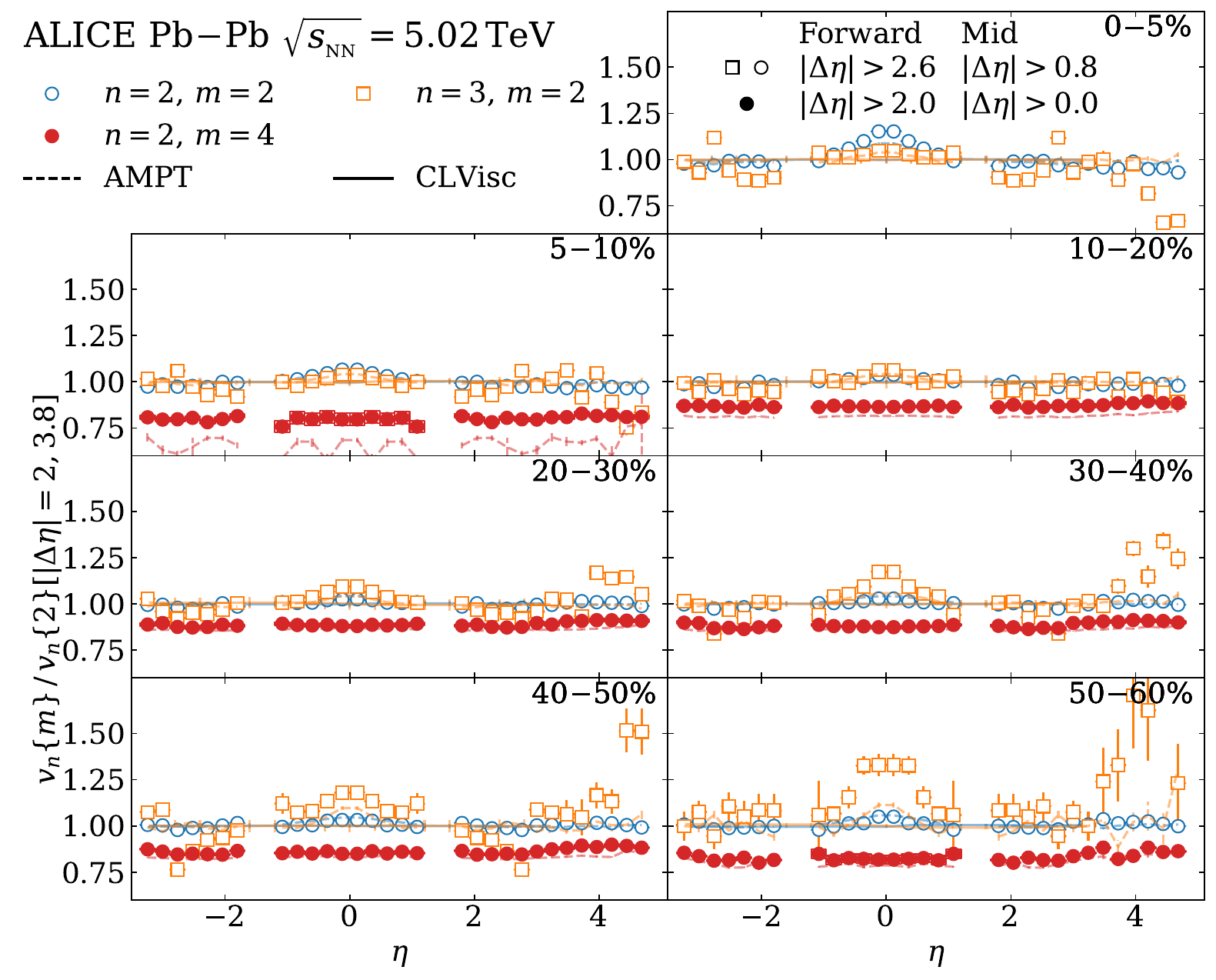}
  \caption{Ratio of two-particle results with a medium-sized
    pseudorapidity separation and 4-particle results to the
    two-particle results employing a large pseudorapidity separation
    between particles of interest and reference particles in \PbPb{}
    collisions at $\sNN{} = 5.02\,\TeV{}$.  Also shown are the same
    ratios for the \AMPT{} and \Hydro{} models. For the latter, the
    ratios are compatible with unity as the \Hydro{} model produces
    negligible non-flow~\cite{Wu:2018cpc}.}
  \label{fig:result:pbpb:ratio}
\end{figure}

To better appreciate the difference between the results shown in
Figs.~\ref{fig:result:pbpb:gap08} and~\ref{fig:result:pbpb:gap20},
\FigRef{fig:result:pbpb:ratio} presents the ratio of $\vnm{n}{2}$ with
reference region at midrapidity, and the $\vnm{2}{4}$ results, to the
$\vnm{n}{2}$ results using the forward reference region.  The relevant
ratio for $\vnm{4}{2}$ has large statistical uncertainties and is
therefore not shown. The fluctuation of data points at large pseudorapidity,
in particular for larger centralities, are remnants of systematic effects
which have not been identified and therefore not quantified,
and are further accentuated when calculating the ratio. 
The ratio exhibits no significant dependence on
the pseudorapidity, except near $\eta=0$.  This qualitative difference
only at midrapidity could be understood as a result of better
suppression of short-range correlations, or non-flow, when a larger
\etagap{} is used.  At forward pseudorapidity, where the \etagap{} is
large, the contribution from such correlations may be already
sufficiently suppressed, as the data suggest.

The clear difference between the values of $v_2$ using two- and
four-particle cumulants, reflected by their ratio being significantly
smaller than unity, can be mainly attributed to opposite contributions
of the event-by-event fluctuations of the flow probability density
function, in particular its variance, to different order
cumulants~\cite{Voloshin:2007pc,Yan:2013laa,Yan:2014afa,Gronqvist:2016hym},
and partly also to better suppression of non-flow correlations in case
of $v_2\{4\}$.  The ratio $\vnm{2}{4}/\vnm{2}{2}$ is assumed to
reflect the ratio $\varepsilon_2\{4\}/\varepsilon_2\{2\}$, since the
second order flow magnitude $\vn{2}$ is proportional to the second
order eccentricity $\varepsilon_2$ of the initial overlap
region~\cite{Holopainen:2010gz,Qin:2010pf,Qiu:2011iv,Gale:2012rq,Niemi:2012aj}.
It has previously been reported for the integrated flow measurements
that this ratio exhibits a deviation from unity. This deviation is
larger in more central collisions, affected by fluctuations in the
initial spatial
asymmetry~\cite{Acharya:2018lmh,ALICE:2018lao,CMS:2019cyz}, and shows
potential to constrain the different initial state models. Our
measurements of pseudorapidity differential ratio of
$\vnm{2}{4}/\vnm{2}{2}$, shown in \FigRef{fig:result:pbpb:ratio},
provide more detailed understanding of flow fluctuations. Similar
dependence on centrality as the integrated
measurements~\cite{Acharya:2018lmh} is reported, but the results
indicate that the variance of $\vn{n}$, thus also the variance of
$\epsilon_n$ in the initial state, is invariant with pseudorapidity.
The same observation was reported in Ref.~\cite{Denicol:2015nhu} based
on hydrodynamic calculations.  The authors confirmed that the
event-by-event flow fluctuations are only weakly sensitive to
pseudorapidity, and are close to the relative eccentricity
fluctuations over a wide rapidity range.  Therefore, ratios of
cumulants of $v_n$ distributions presented here as a function of
pseudorapidity provide important input to constraining fluctuations of
the 3D initial state~\cite{Denicol:2015nhu}.

Comparisons with calculations from the AMPT model~\cite{Ma:2016fve}
and hydrodynamic model CLVisc~\cite{Wu:2018cpc} are shown in
Figs.~\ref{fig:result:pbpb:gap08} and~\ref{fig:result:pbpb:gap20} with
dashed and full lines, respectively. The \AMPT{} is a hybrid model
that evolves fluctuating initial conditions from the HIJING
model~\cite{Gyulassy:1994ew}, followed by partonic and hadronic
interactions. It gives reasonable descriptions of the rapidity
distributions measured in heavy-ion collisions. The \Hydro{} is a
($3+1$)-dimensional hydrodynamic model that simulates the dynamic
evolution of the QGP fireball based on the initial conditions computed
with the \AMPT{} model. Because of extra parameters tuning and
additional parton smearing in the \Hydro{} to fit the experimental
data, the initial conditions from \Hydro{} are not identical to those
of the \AMPT{} model.  The \AMPT{} and \Hydro{} calculations of the
$\vn{n}$ coefficients shown in Fig.~\ref{fig:result:pbpb:gap08}
and~\ref{fig:result:pbpb:gap20} were carried out in a similar way as
the analysis of ALICE data presented in this article, using the
pseudorapidity ranges and \etagap{} as illustrated in
Fig.~\ref{fig:cumulants:method1}.

Both \AMPT{} and \Hydro{} calculations qualitatively follow the trend
of the pseudorapidity dependence of $\vn{n}$ found in data. Both
models, however, overestimate the $\vn{n}$ coefficients over the whole
presented pseudorapidity range in the $30\%$ most central \PbPb{}
collisions. In the case of \AMPT{}, that trend continues down to the
most peripheral collisions, while \Hydro{} underestimates the
experimental results for the 40--60$\%$ collision centrality range.
The AMPT calculations with a small \etagap{}
(Fig.~\ref{fig:result:pbpb:gap08}) show peaked pseudorapidity
dependence near $\eta=0$, while this trend vanishes with large
\etagap{} (Fig.~\ref{fig:result:pbpb:gap20}), similarly as in data.
In contrast, the \Hydro{} calculations remain unchanged for the two
cases.  This further substantiates the origin of this peak to be
caused by non-flow contributions from short range correlations, since
\Hydro{} produces less of these than the transport \AMPT{} model.
Nevertheless, comparison with measurements with forward reference
regions (i.e. large \etagap{}) in Fig.~\ref{fig:result:pbpb:gap20}
reveals that the \Hydro{} model predicts a systematically more peaked
distribution near midrapidity, while \AMPT{} shows a constant $\eta$
dependence of $v_n$, which is in qualitative agreement with the
data. Other calculations of hydrodynamical
models~\cite{Wu:2018cpc,Denicol:2015nhu,Molnar:2014zha} attempted to
describe the prior datasets, with no quantitative agreement being
reached so far. This suggests, that the longitudinal structure of the
initial state or longitudinal evolution of the system are not yet
properly understood in these models and more theoretical
investigations are warranted.

\subsection{Flow measurements in \XeXe{} collisions}

\paragraph{Modest pseudorapidity separation}

\begin{figure}[htbp]
  \centering
  \includegraphics[width=\figwidth]{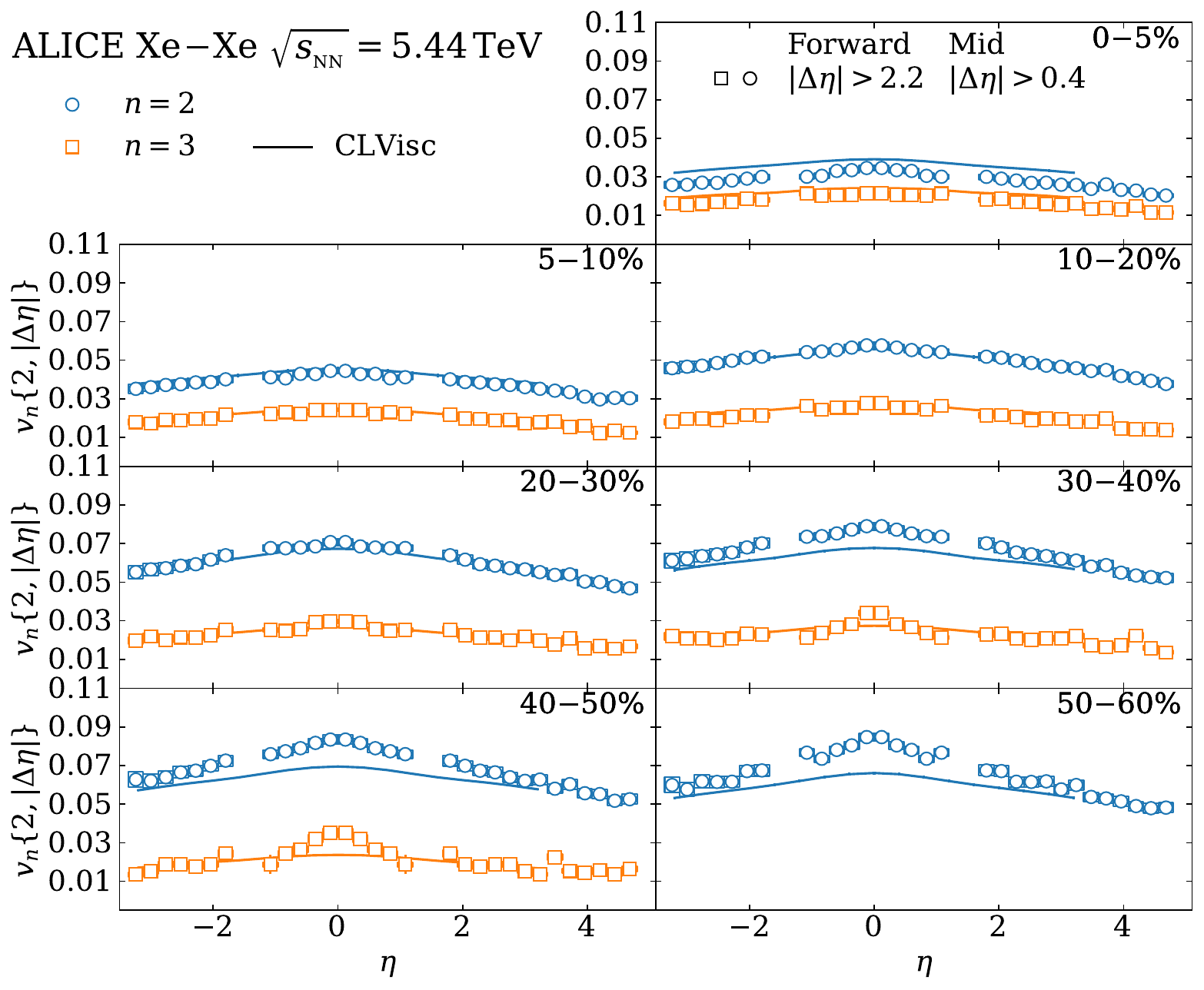}
  \caption[$\vnm{n}{2,|\Delta \eta| > 0.4}$ with TPC reference in
  \XeXe{} collisions.]{The differential flow measurements,
    $\vnm{n}{2}$, measured with the 2-particle cumulant and choosing
    reference particles from the TPC in \XeXe{} collisions at
    $\sNN = 5.44\,\TeV{}$. The choice of the reference particles
    results in pseudorapidity separations of $|\Delta \eta | > 0.4$
    and $2.2$ at mid and forward rapidities, respectively.  Also shown
    are results from the \Hydro{} model.}
  \label{fig:result:xexe:gap04}
\end{figure}

\begin{figure}[htbp]
  \centering
  \includegraphics[width=\figwidth]{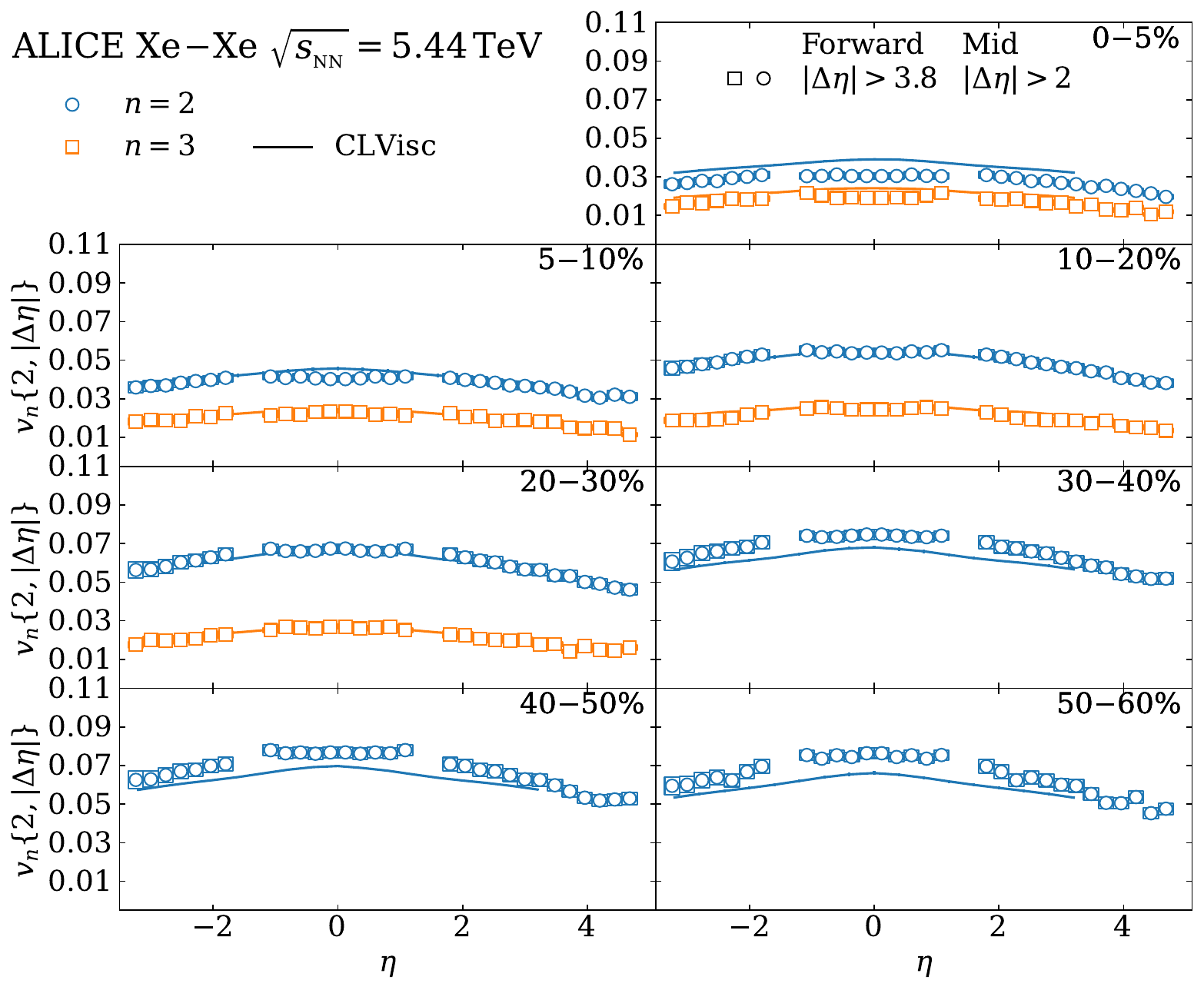}
  \caption[$\vnm{n}{2,|\Delta \eta| > 2.0}$ with FMD reference in
  \XeXe{} collisions.]{The differential flow measurements,
    $\vnm{n}{2}$, measured with the 2-particle cumulant and choosing
    reference particles chosen from the FMD in \XeXe{} collisions at
    $\sNN{} = 5.02\,\TeV{}$, with pseudorapidity separations of
    $|\Delta \eta | > 2.0$ and $3.8$ at mid and forward rapidities,
    respectively.  Also shown are results from the \Hydro{} model.}
  \label{fig:result:xexe:gap20}
\end{figure}

\begin{figure}[htbp]
  \centering
  \includegraphics[width=\figwidth]{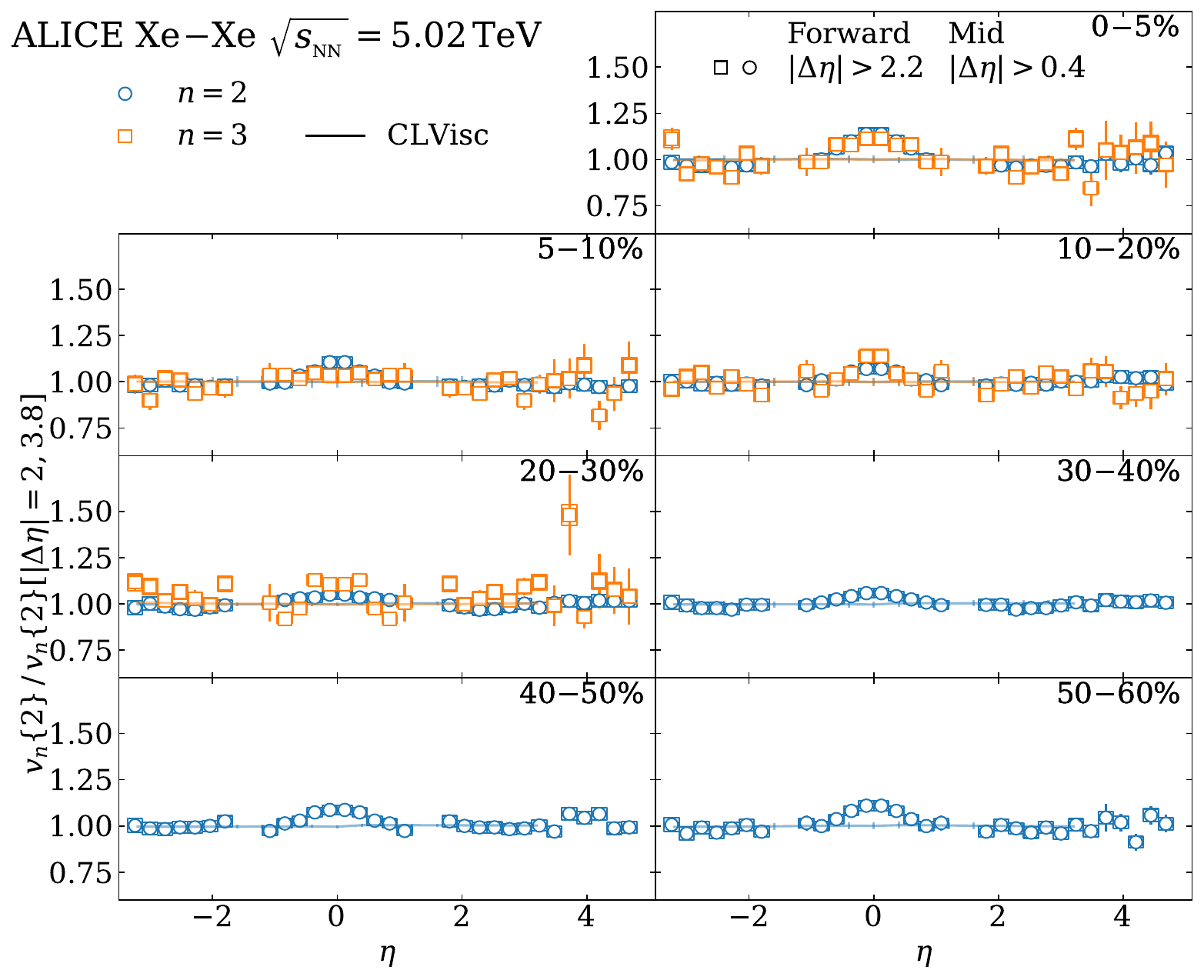}
  \caption{Ratio of two-particle results with a medium-sized
    separation to the two-particle results employing a large
    separation between particles of interest and reference
    particles in \XeXe{} collisions at $\sNN{} = 5.44\,\TeV{}$.
    Also shown are the same ratios for the \AMPT{} and \Hydro{} models. For the latter, 
    the ratios are compatible with unity as the \Hydro{} model produces negligible non-flow~\cite{Wu:2018cpc}.}
\label{fig:result:xexe:ratio}
\end{figure}

The flow coefficients $v_n\{2\}$ in \XeXe{} collisions at
$\sNN=~5.44\,\TeV{}$ are shown as a function of pseudorapidity and
centrality in \FigRef{fig:result:xexe:gap04} for the case where the
reference region is at midrapidity (i.e a modest \etagap{}), and in
\FigRef{fig:result:xexe:gap20} when the reference region resides at
forward rapidity (i.e. a large \etagap).  Due to limited amount of
data available from \XeXe{} collisions, the choice of \etagap{}
separation for $\vnm{2}{2}$ and $\vnm{3}{2}$ was decreased down to
$|\Delta \eta| > 0.4$ at midrapidity and to $2.2$ at forward rapidity,
and neither the $\vnm{4}{2}$ nor $\vnm{2}{4}$ could be measured in
these collisions.  The results from \XeXe{} collisions show a similar
trend as seen in \PbPb{} collisions with roughly $30\%$ larger
magnitude in the $5\%$ most central collisions, which was explained by
larger deformation of the xenon nucleus~\cite{ALICE:2018lao}.
Similarly to \PbPb{} collisions, measurements with \RO{} from a
forward $\eta$ region lead to a less pronounced dependence on
pseudorapidity near midrapidity due to a larger \etagap{} between the
correlated \RO{} and \OIO{}.  This is further illustrated in
\FigRef{fig:result:xexe:ratio} by the ratio of $v_n$ coefficients
obtained with different reference regions, thus different \etagap .

Calculations from the \Hydro{} model~\cite{Wu:2018cpc} are compared
with the experimental results in Figs.~\ref{fig:result:xexe:gap04}
and~\ref{fig:result:xexe:gap20}. The model qualitatively describes the
data, although with a more peaked $\eta$ dependence at midrapidity
compared with the results using a large \etagap{}, and it also
overestimates (underestimates) the $v_n$ measurements at central
(peripheral) collisions. The shift between the two scenarios happens
at higher centrality (around 20\%) as compared with the \PbPb{}
collisions (around 40\%).

\paragraph{decorrelation of the flow vectors}

\begin{figure}[htbp]
  \centering
  \includegraphics[width=\figwidth]{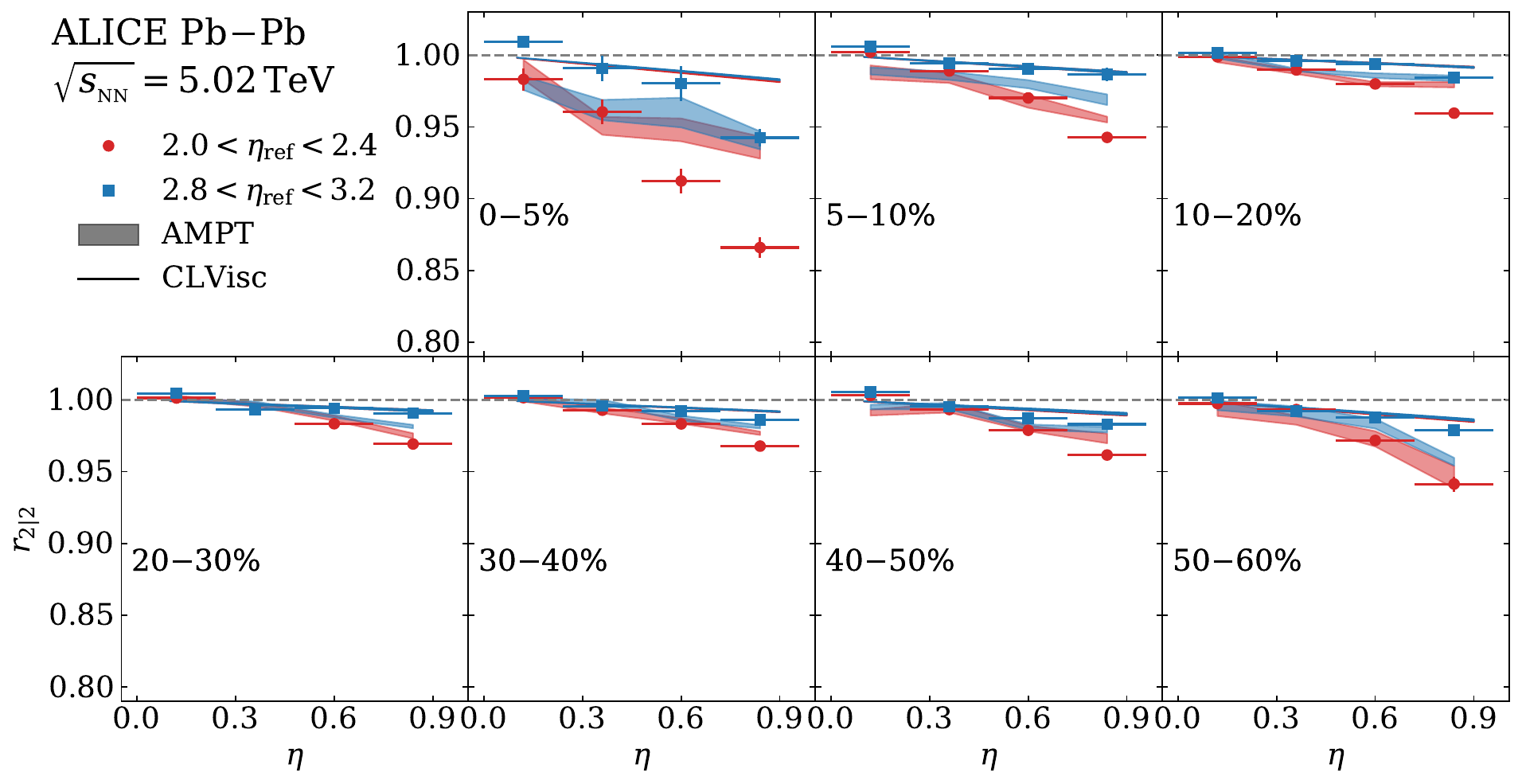}
  \caption[$\rnn{2}$ with different choices for
  $\eta_{\mathrm{ref}}$]{$\rnn{2}$ with different choices for of
    reference region $\eta$ in \PbPb{} collisions at
    $\sNN=5.02\,\TeV{}$. \AMPT{} and \Hydro{} model calculations are
    compared with the results.}
  \label{fig:result:pbpb:r22}
\end{figure}

Measurements of $\rnn{2}$ as a function of absolute pseudorapidity
$\eta$ for different centrality classes of \PbPb{} collisions at
$\sNN=5.02\,\TeV{}$ are presented in \FigRef{fig:result:pbpb:r22}. The
red and blue markers represent two different cases of the absolute
reference pseudorapidity regions, chosen to be
$2 < \eta_{\mathrm{ref}} < 2.4$ or $2.8 < \eta_{\mathrm{ref}} < 3.2$,
respectively (\FigRef{fig:cumulants:method2}(d)).  It can be observed
that the measurements of $\rnn{2}$ generally deviate from unity and
this deviation is stronger in most central and peripheral collisions,
while it is less pronounced in midcentral collisions, in line with
dominant average elliptic geometry at these collision centralities.

Two perspectives of flow vector fluctuations can be studied from
\FigRef{fig:result:pbpb:r22}.  First, for fixed $\eta_{\mathrm{ref}}$,
when $\eta$ increases, the longitudinal decorrelation between two
narrow intervals of \OIO{}, $\eta$ and $-\eta$ is studied.  At
$\eta\sim 0$ (i.e. no relative separation), the correlations in both
numerator and denominator of $\rnn{2}$ have similar longitudinal
separation with respect to $\eta_{\mathrm{ref}}$, therefore any
effects arising from this \etagap{} cancel out in the $\rnn{2}$
ratio. On the contrary, the $\rnn{2}$ measurements at $\eta \sim 1.0$
have a large \etagap{} between correlated hadrons in the numerator
($|-\eta - \eta_{\mathrm{ref}}|$), and a small \etagap{} in the
denominator ($|\eta - \eta_{\mathrm{ref}}|$), which in the presence of
flow vector fluctuations would lead to $\rnn{2} < 1$. The results
presented in \FigRef{fig:result:pbpb:r22} therefore suggest the
presence of longitudinal flow vector fluctuations, in agreement with
prior observations in
Refs.~\cite{CMS:2015xmx,ATLAS:2017rij,ATLAS:2020sgl}.

Secondly, for fixed $\eta$ of the \OIO{}, the evolution of flow vector
fluctuations with pseudorapidity can be addressed by investigating
different choices of $\eta_{\mathrm{ref}}$. If the fluctuations
increase linearly with pseudorapidity, then the decorrelation effect
from the pseudorapidity separation $\eta - \eta_{\mathrm{ref}}$ would
cancel out in $\rnn{2}$, and $\rnn{2}$ will only reflect the
decorrelation between $\eta$ and $-\eta$.  That is, a linear increase
in fluctuations would lead to a $\rnn{2}$ ratio independent of the
choice of $\eta_{\mathrm{ref}}$.  The measurements of $\rnn{2}$ shown
in \FigRef{fig:result:pbpb:r22} however exhibit a significant
difference for different choices of the reference region, with more
distant $\eta_{\mathrm{ref}}$ showing smaller decorrelation.  This
suggests, that the effect of flow fluctuations saturates at a
particular value of \etagap{}.  In case of a distant
$\eta_{\mathrm{ref}}$, the numerator would reach a saturation point,
while the \etagap{} in the denominator is insufficient to reach it. On
the contrary, in case of a close $\eta_{\mathrm{ref}}$, neither the
numerator nor the denominator would reach the saturation point,
resulting in larger deviation of $\rnn{2}$ from unity. Under this
assumption, $\rnn{2}\sim 1$ in the limiting case when both numerator
and denominator would have the \etagap{} large enough for the
fluctuations effect to be saturated. Therefore, the fact that our
results with large $\eta_{\mathrm{ref}}$ exhibit smaller, but still
statistically significant, decorrelation effect, suggests that the
limiting \etagap{} is in the range $|\Delta\eta| > 2$.

A similar dependence on $\eta_{\mathrm{ref}}$ was found in
Refs.~\cite{CMS:2015xmx,ATLAS:2017rij,ATLAS:2020sgl}, where it was
argued that the difference of $\rnn{2}$ from different
$\eta_{\mathrm{ref}}$ configurations may alter the presence of
short-range non-flow correlations, leading to stronger artificial
deviation of $\rnn{2}$ from unity in case $\eta_{\mathrm{ref}}$ is
positioned closer to midrapidity. However, the same analysis in the
HIJING model~\cite{Gyulassy:1994ew}, which does not include any
collective motion (not shown in the article), demonstrated that even
with modest \etagap{}, the non-flow contamination in both numerator
and denominator of $\rnn{n}$ in heavy-ion collisions is negligible. In
addition, non-flow effects tend to be more dominant in peripheral
collisions, in contrast to the observations made in
\FigRef{fig:result:pbpb:r22}, where the largest difference of
$\rnn{2}$ from different $\eta_{\mathrm{ref}}$ configurations was
found in the most central collisions. The difference between the
measurements with respect to the two $\eta_{\mathrm{ref}}$ regions
seen in \FigRef{fig:result:pbpb:r22}, therefore, cannot be explained
solely by non-flow effects, and the deviation of $\rnn{2}$ from unity
hints to sizable longitudinal flow vector fluctuations, which saturate
at large pseudorapidity separations.

\begin{figure}[htbp]
  \centering
  \includegraphics[width=\figwidth]{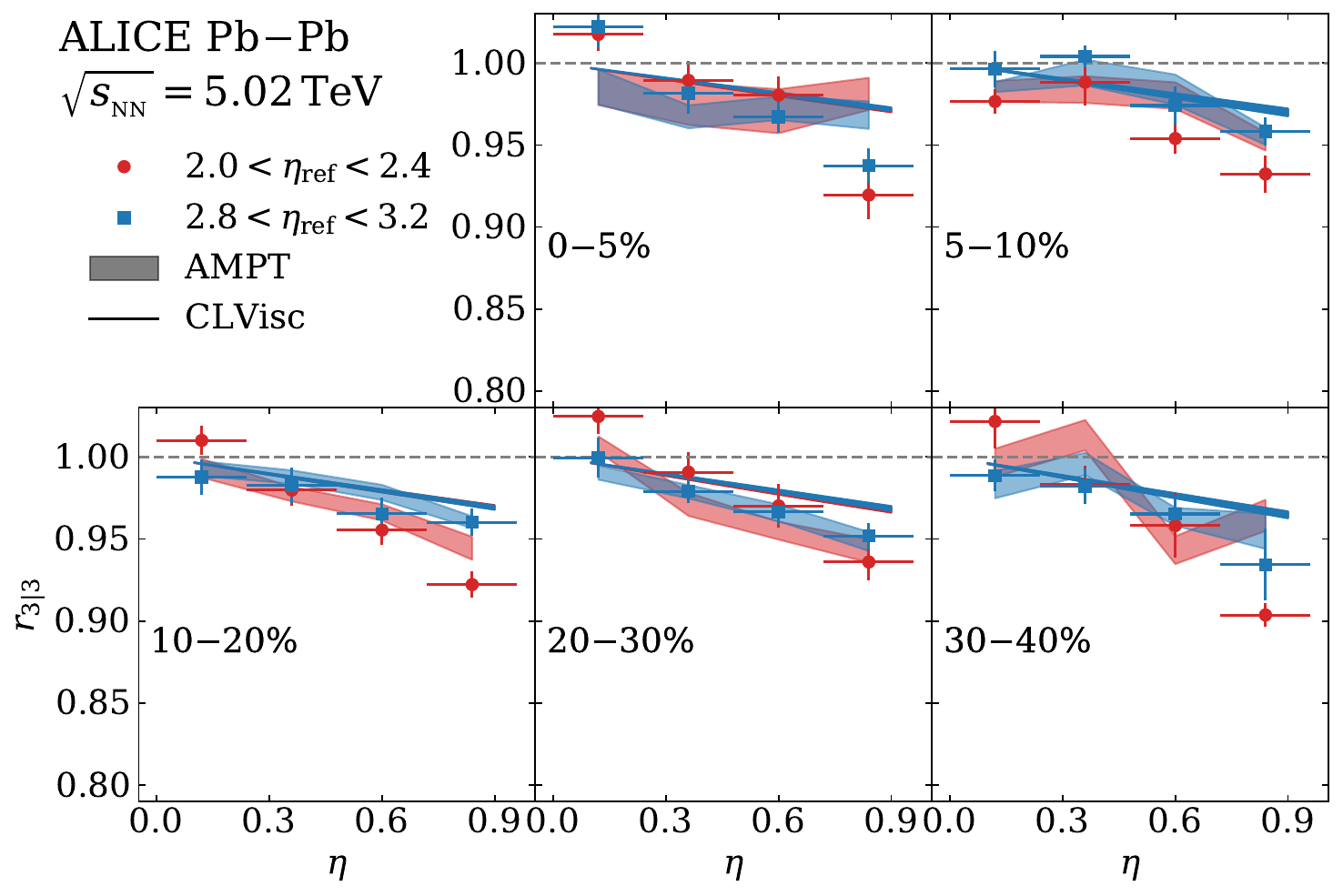}
  \caption[$\rnn{3}$ with different choices for the reference region
  $\eta$.]{$\rnn{3}$ with different choices for reference region
    $\eta$ in \PbPb{} collisions at $\sNN=5.02\,\TeV{}$.  \AMPT{} and
    \Hydro{} model calculations are compared with the results.}
  \label{fig:result:pbpb:r33}
\end{figure}

The measurements of $\rnn{3}$ in \PbPb{} collisions at
$\sNN=5.02\,\TeV{}$ are presented in \FigRef{fig:result:pbpb:r33} as a
function of absolute $\eta$ at different centrality classes, and for
the two choices of the absolute $\eta_{\mathrm{ref}}$ regions. Results
of $\rnn{3}$ for collision centralities larger than 40\% are omitted
due to large statistical uncertainties. Overall, $\rnn{3}$ shows a
clear deviation from unity with larger magnitude than $\rnn{2}$.  This
observation is analogous to results presented in
Ref.~\cite{CMS:2015xmx,ATLAS:2017rij} that included also the
$4^{\rm {th}}$ harmonic, and is in line with expectations from
fluctuation-driven harmonics of higher order. In addition, $\rnn{3}$
exhibits weaker centrality dependence as compared to $\rnn{2}$,
similar to the observations of $\vn{2}$ and $\vn{3}$ coefficients
shown in Figs.~\ref{fig:result:pbpb:gap08} and
\ref{fig:result:pbpb:gap20}. Similarly as in
Ref.~\cite{CMS:2015xmx,ATLAS:2017rij}, measurements with different
ranges of $\eta_{\mathrm{ref}}$ are compatible with each other within
the statistical uncertainties, which suggest different $\eta$
dependence of the longitudinal flow vector fluctuations compared with
the $\rnn{2}$ measurement.

Results presented in Figs.~\ref{fig:result:pbpb:r22} and
\ref{fig:result:pbpb:r33} are compared with the \AMPT{} and \Hydro{}
model calculations.  As opposed to the measurements, the models do not
distinguish between the choices of the $\eta_{\mathrm{ref}}$ regions
used for correlations.  This substantiates the observation that the
deviation from unity is not driven by short range correlations.  In
case of $2.8 < |\eta_{\mathrm{ref}}| < 3.2$, both models reproduce the
trend of the data within uncertainties.  On the contrary, when the
reference region is chosen as $2.0 < |\eta_{\mathrm{ref}}| < 2.4$, the
\Hydro{} significantly underestimates the effect of longitudinal flow
fluctuations, while \AMPT{} reproduces the trend of the data except
for the $5\%$ most central collisions.  Both models are able to
reproduce the magnitude and trend of decorrelation of the $\rnn{3}$
results shown in \FigRef{fig:result:pbpb:r33} within the
uncertainties.

\section{Summary}

The pseudorapidity and centrality dependence of $\vn{2}$, $\vn{3}$,
and $\vn{4}$ in both \PbPb{} at ${\sNN=5.02\,\TeV}$ and \XeXe{} at
${\sNN=5.44\,\TeV}$ is measured over the widest possible longitudinal
range. The results show a smooth pseudorapidity dependence modulated
by a strong collision centrality dependence. Non-flow contributions
are minimised by using a pseudorapidity gap $|\Delta\eta|>2$ or by
using four-particle cumulants. The ratio of the second order flow
coefficient calculated with four-particle cumulant
($\vnm{2}{4,|\Delta\eta|>0}$) to the one obtained with two-particle
cumulant with a large \etagap{} ($\vnm{2}{2,|\Delta\eta|>2,3.8}$) is
almost constant over the whole pseudorapidity range. This result
suggests that the variance of flow probability density function is
independent of pseudorapidity, providing constraints to the models of
fluctuating initial state.  The longitudinal flow vector fluctuations
are investigated via the measurements of the ratio $r_{n|n}$ to
further quantify this observation. The $r_{2|2}$ exhibits a clear
deviation from unity, which is more pronounced in central than
peripheral collisions. The difference between the measured
decorrelation with respect to different $\eta_{\mathrm{ref}}$ regions
points to potential saturation effect of longitudinal flow vector
fluctuations above a certain pseudorapidity separation. This is in
contrast to current understanding based on previous publications at
the LHC.  The $r_{3|3}$ measurements reveal the effect of flow vector
fluctuations of stronger magnitude, but with a weaker centrality
dependence, compared to $r_{2|2}$. This confirms the
fluctuation-driven nature of higher order harmonics.  Both the \AMPT{}
and the \Hydro{} models can qualitatively reproduce the experimental
results of $\vn{n}$, but differ in important details such as overall
amplitude and exact pseudorapidity dependence, especially near
midrapidity. The measurements of $r_{n|n}$ are qualitatively
reproduced by both models, except for the dependence on the choice of
the reference region $\eta_{\mathrm{ref}}$ of $r_{2|2}$, which was not
observed in the model calculations.  These findings reveal that the
current state-of-the-art models are not able to simultaneously
describe the pseudorapidity dependence of anisotropic flow and its
fluctuations. Our results highlight the inadequacies in the current
understanding of particle production, particularly in the longitudinal
direction, and shall help to improve the modelling of longitudinal
fluctuations of initial conditions and the subsequent system
evolution.


\newenvironment{acknowledgement}{\relax}{\relax}%
\begin{acknowledgement}%
\section*{Acknowledgements}

The ALICE Collaboration would like to thank all its engineers and technicians for their invaluable contributions to the construction of the experiment and the CERN accelerator teams for the outstanding performance of the LHC complex.
The ALICE Collaboration gratefully acknowledges the resources and support provided by all Grid centres and the Worldwide LHC Computing Grid (WLCG) collaboration.
The ALICE Collaboration acknowledges the following funding agencies for their support in building and running the ALICE detector:
A. I. Alikhanyan National Science Laboratory (Yerevan Physics Institute) Foundation (ANSL), State Committee of Science and World Federation of Scientists (WFS), Armenia;
Austrian Academy of Sciences, Austrian Science Fund (FWF): [M 2467-N36] and Nationalstiftung f\"{u}r Forschung, Technologie und Entwicklung, Austria;
Ministry of Communications and High Technologies, National Nuclear Research Center, Azerbaijan;
Conselho Nacional de Desenvolvimento Cient\'{\i}fico e Tecnol\'{o}gico (CNPq), Financiadora de Estudos e Projetos (Finep), Funda\c{c}\~{a}o de Amparo \`{a} Pesquisa do Estado de S\~{a}o Paulo (FAPESP) and Universidade Federal do Rio Grande do Sul (UFRGS), Brazil;
Bulgarian Ministry of Education and Science, within the National Roadmap for Research Infrastructures 2020-2027 (object CERN), Bulgaria;
Ministry of Education of China (MOEC) , Ministry of Science \& Technology of China (MSTC) and National Natural Science Foundation of China (NSFC), China;
Ministry of Science and Education and Croatian Science Foundation, Croatia;
Centro de Aplicaciones Tecnol\'{o}gicas y Desarrollo Nuclear (CEADEN), Cubaenerg\'{\i}a, Cuba;
Ministry of Education, Youth and Sports of the Czech Republic, Czech Republic;
The Danish Council for Independent Research | Natural Sciences, the VILLUM FONDEN and Danish National Research Foundation (DNRF), Denmark;
Helsinki Institute of Physics (HIP), Finland;
Commissariat \`{a} l'Energie Atomique (CEA) and Institut National de Physique Nucl\'{e}aire et de Physique des Particules (IN2P3) and Centre National de la Recherche Scientifique (CNRS), France;
Bundesministerium f\"{u}r Bildung und Forschung (BMBF) and GSI Helmholtzzentrum f\"{u}r Schwerionenforschung GmbH, Germany;
General Secretariat for Research and Technology, Ministry of Education, Research and Religions, Greece;
National Research, Development and Innovation Office, Hungary;
Department of Atomic Energy Government of India (DAE), Department of Science and Technology, Government of India (DST), University Grants Commission, Government of India (UGC) and Council of Scientific and Industrial Research (CSIR), India;
National Research and Innovation Agency - BRIN, Indonesia;
Istituto Nazionale di Fisica Nucleare (INFN), Italy;
Japanese Ministry of Education, Culture, Sports, Science and Technology (MEXT) and Japan Society for the Promotion of Science (JSPS) KAKENHI, Japan;
Consejo Nacional de Ciencia (CONACYT) y Tecnolog\'{i}a, through Fondo de Cooperaci\'{o}n Internacional en Ciencia y Tecnolog\'{i}a (FONCICYT) and Direcci\'{o}n General de Asuntos del Personal Academico (DGAPA), Mexico;
Nederlandse Organisatie voor Wetenschappelijk Onderzoek (NWO), Netherlands;
The Research Council of Norway, Norway;
Commission on Science and Technology for Sustainable Development in the South (COMSATS), Pakistan;
Pontificia Universidad Cat\'{o}lica del Per\'{u}, Peru;
Ministry of Education and Science, National Science Centre and WUT ID-UB, Poland;
Korea Institute of Science and Technology Information and National Research Foundation of Korea (NRF), Republic of Korea;
Ministry of Education and Scientific Research, Institute of Atomic Physics, Ministry of Research and Innovation and Institute of Atomic Physics and Universitatea Nationala de Stiinta si Tehnologie Politehnica Bucuresti, Romania
Ministry of Education, Science, Research and Sport of the Slovak Republic, Slovakia;
National Research Foundation of South Africa, South Africa;
Swedish Research Council (VR) and Knut \& Alice Wallenberg Foundation (KAW), Sweden;
European Organization for Nuclear Research, Switzerland;
Suranaree University of Technology (SUT), National Science and Technology Development Agency (NSTDA) and National Science, Research and Innovation Fund (NSRF via PMU-B B05F650021), Thailand;
Turkish Energy, Nuclear and Mineral Research Agency (TENMAK), Turkey;
National Academy of  Sciences of Ukraine, Ukraine;
Science and Technology Facilities Council (STFC), United Kingdom;
National Science Foundation of the United States of America (NSF) and United States Department of Energy, Office of Nuclear Physics (DOE NP), United States of America.
In addition, individual groups or members have received support from:
European Research Council, Strong 2020 - Horizon 2020 (grant nos. 950692, 824093), European Union;
Academy of Finland (Center of Excellence in Quark Matter) (grant nos. 346327, 346328), Finland.

\end{acknowledgement}

\bibliographystyle{utphys}

\bibliography{other}

\newpage
\appendix

%
%

\section{The ALICE Collaboration}
\label{app:collab}
\begin{flushleft} 
\small

S.~Acharya\,\orcidlink{0000-0002-9213-5329}\,$^{\rm 128}$, 
D.~Adamov\'{a}\,\orcidlink{0000-0002-0504-7428}\,$^{\rm 87}$, 
A.~Adler$^{\rm 71}$, 
G.~Aglieri Rinella\,\orcidlink{0000-0002-9611-3696}\,$^{\rm 33}$, 
M.~Agnello\,\orcidlink{0000-0002-0760-5075}\,$^{\rm 30}$, 
N.~Agrawal\,\orcidlink{0000-0003-0348-9836}\,$^{\rm 52}$, 
Z.~Ahammed\,\orcidlink{0000-0001-5241-7412}\,$^{\rm 136}$, 
S.~Ahmad\,\orcidlink{0000-0003-0497-5705}\,$^{\rm 16}$, 
S.U.~Ahn\,\orcidlink{0000-0001-8847-489X}\,$^{\rm 72}$, 
I.~Ahuja\,\orcidlink{0000-0002-4417-1392}\,$^{\rm 38}$
% ...

\section*{Affiliation Notes}

$^{\rm I}$ Deceased\\
$^{\rm II}$ Also at: Max-Planck-Institut fur Physik, Munich, Germany\\
$^{\rm III}$ Also at: Italian National Agency for New Technologies, Energy and Sustainable Economic Development (ENEA), Bologna, Italy\\
$^{\rm IV}$ Also at: Dipartimento DET del Politecnico di Torino, Turin, Italy\\
$^{\rm V}$ Also at: Department of Applied Physics, Aligarh Muslim University, Aligarh, India\\
$^{\rm VI}$ Also at: Institute of Theoretical Physics, University of Wroclaw, Poland\\
$^{\rm VII}$ Also at: An institution covered by a cooperation agreement with CERN\\

\section*{Collaboration Institutes}

$^{1}$ A.I. Alikhanyan National Science Laboratory (Yerevan Physics Institute) Foundation, Yerevan, Armenia\\
$^{2}$ AGH University of Krakow, Cracow, Poland\\
$^{3}$ Bogolyubov Institute for Theoretical Physics, National Academy of Sciences of Ukraine, Kiev, Ukraine\\
$^{4}$ Bose Institute, Department of Physics  and Centre for Astroparticle Physics and Space Science (CAPSS), Kolkata, India\\
$^{5}$ California Polytechnic State University, San Luis Obispo, California, United States\\
$^{6}$ Central China Normal University, Wuhan, China\\
$^{7}$ Centro de Aplicaciones Tecnol\'{o}gicas y Desarrollo Nuclear (CEADEN), Havana, Cuba\\
$^{8}$ Centro de Investigaci\'{o}n y de Estudios Avanzados (CINVESTAV), Mexico City and M\'{e}rida, Mexico\\
$^{9}$ Chicago State University, Chicago, Illinois, United States\\
$^{10}$ China Institute of Atomic Energy, Beijing, China\\
$^{11}$ China University of Geosciences, Wuhan, China\\
$^{12}$ Chungbuk National University, Cheongju, Republic of Korea\\
$^{13}$ Comenius University Bratislava, Faculty of Mathematics, Physics and Informatics, Bratislava, Slovak Republic\\
$^{14}$ COMSATS University Islamabad, Islamabad, Pakistan\\
$^{15}$ Creighton University, Omaha, Nebraska, United States\\
$^{16}$ Department of Physics, Aligarh Muslim University, Aligarh, India\\
$^{17}$ Department of Physics, Pusan National University, Pusan, Republic of Korea\\
$^{18}$ Department of Physics, Sejong University, Seoul, Republic of Korea\\
$^{19}$ Department of Physics, University of California, Berkeley, California, United States\\
$^{20}$ Department of Physics, University of Oslo, Oslo, Norway\\
$^{21}$ Department of Physics and Technology, University of Bergen, Bergen, Norway\\
$^{22}$ Dipartimento di Fisica, Universit\`{a} di Pavia, Pavia, Italy\\
$^{23}$ Dipartimento di Fisica dell'Universit\`{a} and Sezione INFN, Cagliari, Italy\\
$^{24}$ Dipartimento di Fisica dell'Universit\`{a} and Sezione INFN, Trieste, Italy\\
$^{25}$ Dipartimento di Fisica dell'Universit\`{a} and Sezione INFN, Turin, Italy\\
$^{26}$ Dipartimento di Fisica e Astronomia dell'Universit\`{a} and Sezione INFN, Bologna, Italy\\
$^{27}$ Dipartimento di Fisica e Astronomia dell'Universit\`{a} and Sezione INFN, Catania, Italy\\
$^{28}$ Dipartimento di Fisica e Astronomia dell'Universit\`{a} and Sezione INFN, Padova, Italy\\
$^{29}$ Dipartimento di Fisica `E.R.~Caianiello' dell'Universit\`{a} and Gruppo Collegato INFN, Salerno, Italy\\
$^{30}$ Dipartimento DISAT del Politecnico and Sezione INFN, Turin, Italy\\
$^{31}$ Dipartimento di Scienze MIFT, Universit\`{a} di Messina, Messina, Italy\\
$^{32}$ Dipartimento Interateneo di Fisica `M.~Merlin' and Sezione INFN, Bari, Italy\\
$^{33}$ European Organization for Nuclear Research (CERN), Geneva, Switzerland\\
$^{34}$ Faculty of Electrical Engineering, Mechanical Engineering and Naval Architecture, University of Split, Split, Croatia\\
$^{35}$ Faculty of Engineering and Science, Western Norway University of Applied Sciences, Bergen, Norway\\
$^{36}$ Faculty of Nuclear Sciences and Physical Engineering, Czech Technical University in Prague, Prague, Czech Republic\\
$^{37}$ Faculty of Physics, Sofia University, Sofia, Bulgaria\\
$^{38}$ Faculty of Science, P.J.~\v{S}af\'{a}rik University, Ko\v{s}ice, Slovak Republic\\
$^{39}$ Frankfurt Institute for Advanced Studies, Johann Wolfgang Goethe-Universit\"{a}t Frankfurt, Frankfurt, Germany\\
$^{40}$ Fudan University, Shanghai, China\\
$^{41}$ Gangneung-Wonju National University, Gangneung, Republic of Korea\\
$^{42}$ Gauhati University, Department of Physics, Guwahati, India\\
$^{43}$ Helmholtz-Institut f\"{u}r Strahlen- und Kernphysik, Rheinische Friedrich-Wilhelms-Universit\"{a}t Bonn, Bonn, Germany\\
$^{44}$ Helsinki Institute of Physics (HIP), Helsinki, Finland\\
$^{45}$ High Energy Physics Group,  Universidad Aut\'{o}noma de Puebla, Puebla, Mexico\\
$^{46}$ Horia Hulubei National Institute of Physics and Nuclear Engineering, Bucharest, Romania\\
$^{47}$ HUN-REN Wigner Research Centre for Physics, Budapest, Hungary\\
$^{48}$ Indian Institute of Technology Bombay (IIT), Mumbai, India\\
$^{49}$ Indian Institute of Technology Indore, Indore, India\\
$^{50}$ INFN, Laboratori Nazionali di Frascati, Frascati, Italy\\
$^{51}$ INFN, Sezione di Bari, Bari, Italy\\
$^{52}$ INFN, Sezione di Bologna, Bologna, Italy\\
$^{53}$ INFN, Sezione di Cagliari, Cagliari, Italy\\
$^{54}$ INFN, Sezione di Catania, Catania, Italy\\
$^{55}$ INFN, Sezione di Padova, Padova, Italy\\
$^{56}$ INFN, Sezione di Pavia, Pavia, Italy\\
$^{57}$ INFN, Sezione di Torino, Turin, Italy\\
$^{58}$ INFN, Sezione di Trieste, Trieste, Italy\\
$^{59}$ Inha University, Incheon, Republic of Korea\\
$^{60}$ Institute for Gravitational and Subatomic Physics (GRASP), Utrecht University/Nikhef, Utrecht, Netherlands\\
$^{61}$ Institute of Experimental Physics, Slovak Academy of Sciences, Ko\v{s}ice, Slovak Republic\\
$^{62}$ Institute of Physics, Homi Bhabha National Institute, Bhubaneswar, India\\
$^{63}$ Institute of Physics of the Czech Academy of Sciences, Prague, Czech Republic\\
$^{64}$ Institute of Space Science (ISS), Bucharest, Romania\\
$^{65}$ Institut f\"{u}r Kernphysik, Johann Wolfgang Goethe-Universit\"{a}t Frankfurt, Frankfurt, Germany\\
$^{66}$ Instituto de Ciencias Nucleares, Universidad Nacional Aut\'{o}noma de M\'{e}xico, Mexico City, Mexico\\
$^{67}$ Instituto de F\'{i}sica, Universidade Federal do Rio Grande do Sul (UFRGS), Porto Alegre, Brazil\\
$^{68}$ Instituto de F\'{\i}sica, Universidad Nacional Aut\'{o}noma de M\'{e}xico, Mexico City, Mexico\\
$^{69}$ iThemba LABS, National Research Foundation, Somerset West, South Africa\\
$^{70}$ Jeonbuk National University, Jeonju, Republic of Korea\\
$^{71}$ Johann-Wolfgang-Goethe Universit\"{a}t Frankfurt Institut f\"{u}r Informatik, Fachbereich Informatik und Mathematik, Frankfurt, Germany\\
$^{72}$ Korea Institute of Science and Technology Information, Daejeon, Republic of Korea\\
$^{73}$ KTO Karatay University, Konya, Turkey\\
$^{74}$ Laboratoire de Physique Subatomique et de Cosmologie, Universit\'{e} Grenoble-Alpes, CNRS-IN2P3, Grenoble, France\\
$^{75}$ Lawrence Berkeley National Laboratory, Berkeley, California, United States\\
$^{76}$ Lund University Department of Physics, Division of Particle Physics, Lund, Sweden\\
$^{77}$ Nagasaki Institute of Applied Science, Nagasaki, Japan\\
$^{78}$ Nara Women{'}s University (NWU), Nara, Japan\\
$^{79}$ National and Kapodistrian University of Athens, School of Science, Department of Physics , Athens, Greece\\
$^{80}$ National Centre for Nuclear Research, Warsaw, Poland\\
$^{81}$ National Institute of Science Education and Research, Homi Bhabha National Institute, Jatni, India\\
$^{82}$ National Nuclear Research Center, Baku, Azerbaijan\\
$^{83}$ National Research and Innovation Agency - BRIN, Jakarta, Indonesia\\
$^{84}$ Niels Bohr Institute, University of Copenhagen, Copenhagen, Denmark\\
$^{85}$ Nikhef, National institute for subatomic physics, Amsterdam, Netherlands\\
$^{86}$ Nuclear Physics Group, STFC Daresbury Laboratory, Daresbury, United Kingdom\\
$^{87}$ Nuclear Physics Institute of the Czech Academy of Sciences, Husinec-\v{R}e\v{z}, Czech Republic\\
$^{88}$ Oak Ridge National Laboratory, Oak Ridge, Tennessee, United States\\
$^{89}$ Ohio State University, Columbus, Ohio, United States\\
$^{90}$ Physics department, Faculty of science, University of Zagreb, Zagreb, Croatia\\
$^{91}$ Physics Department, Panjab University, Chandigarh, India\\
$^{92}$ Physics Department, University of Jammu, Jammu, India\\
$^{93}$ Physics Program and International Institute for Sustainability with Knotted Chiral Meta Matter (SKCM2), Hiroshima University, Hiroshima, Japan\\
$^{94}$ Physikalisches Institut, Eberhard-Karls-Universit\"{a}t T\"{u}bingen, T\"{u}bingen, Germany\\
$^{95}$ Physikalisches Institut, Ruprecht-Karls-Universit\"{a}t Heidelberg, Heidelberg, Germany\\
$^{96}$ Physik Department, Technische Universit\"{a}t M\"{u}nchen, Munich, Germany\\
$^{97}$ Politecnico di Bari and Sezione INFN, Bari, Italy\\
$^{98}$ Research Division and ExtreMe Matter Institute EMMI, GSI Helmholtzzentrum f\"ur Schwerionenforschung GmbH, Darmstadt, Germany\\
$^{99}$ Saga University, Saga, Japan\\
$^{100}$ Saha Institute of Nuclear Physics, Homi Bhabha National Institute, Kolkata, India\\
$^{101}$ School of Physics and Astronomy, University of Birmingham, Birmingham, United Kingdom\\
$^{102}$ Secci\'{o}n F\'{\i}sica, Departamento de Ciencias, Pontificia Universidad Cat\'{o}lica del Per\'{u}, Lima, Peru\\
$^{103}$ Stefan Meyer Institut f\"{u}r Subatomare Physik (SMI), Vienna, Austria\\
$^{104}$ SUBATECH, IMT Atlantique, Nantes Universit\'{e}, CNRS-IN2P3, Nantes, France\\
$^{105}$ Sungkyunkwan University, Suwon City, Republic of Korea\\
$^{106}$ Suranaree University of Technology, Nakhon Ratchasima, Thailand\\
$^{107}$ Technical University of Ko\v{s}ice, Ko\v{s}ice, Slovak Republic\\
$^{108}$ The Henryk Niewodniczanski Institute of Nuclear Physics, Polish Academy of Sciences, Cracow, Poland\\
$^{109}$ The University of Texas at Austin, Austin, Texas, United States\\
$^{110}$ Universidad Aut\'{o}noma de Sinaloa, Culiac\'{a}n, Mexico\\
$^{111}$ Universidade de S\~{a}o Paulo (USP), S\~{a}o Paulo, Brazil\\
$^{112}$ Universidade Estadual de Campinas (UNICAMP), Campinas, Brazil\\
$^{113}$ Universidade Federal do ABC, Santo Andre, Brazil\\
$^{114}$ Universitatea Nationala de Stiinta si Tehnologie Politehnica Bucuresti, Bucharest, Romania\\
$^{115}$ University of Cape Town, Cape Town, South Africa\\
$^{116}$ University of Derby, Derby, United Kingdom\\
$^{117}$ University of Houston, Houston, Texas, United States\\
$^{118}$ University of Jyv\"{a}skyl\"{a}, Jyv\"{a}skyl\"{a}, Finland\\
$^{119}$ University of Kansas, Lawrence, Kansas, United States\\
$^{120}$ University of Liverpool, Liverpool, United Kingdom\\
$^{121}$ University of Science and Technology of China, Hefei, China\\
$^{122}$ University of South-Eastern Norway, Kongsberg, Norway\\
$^{123}$ University of Tennessee, Knoxville, Tennessee, United States\\
$^{124}$ University of the Witwatersrand, Johannesburg, South Africa\\
$^{125}$ University of Tokyo, Tokyo, Japan\\
$^{126}$ University of Tsukuba, Tsukuba, Japan\\
$^{127}$ Universit\"{a}t M\"{u}nster, Institut f\"{u}r Kernphysik, M\"{u}nster, Germany\\
$^{128}$ Universit\'{e} Clermont Auvergne, CNRS/IN2P3, LPC, Clermont-Ferrand, France\\
$^{129}$ Universit\'{e} de Lyon, CNRS/IN2P3, Institut de Physique des 2 Infinis de Lyon, Lyon, France\\
$^{130}$ Universit\'{e} de Strasbourg, CNRS, IPHC UMR 7178, F-67000 Strasbourg, France, Strasbourg, France\\
$^{131}$ Universit\'{e} Paris-Saclay, Centre d'Etudes de Saclay (CEA), IRFU, D\'{e}partment de Physique Nucl\'{e}aire (DPhN), Saclay, France\\
$^{132}$ Universit\'{e}  Paris-Saclay, CNRS/IN2P3, IJCLab, Orsay, France\\
$^{133}$ Universit\`{a} degli Studi di Foggia, Foggia, Italy\\
$^{134}$ Universit\`{a} del Piemonte Orientale, Vercelli, Italy\\
$^{135}$ Universit\`{a} di Brescia, Brescia, Italy\\
$^{136}$ Variable Energy Cyclotron Centre, Homi Bhabha National Institute, Kolkata, India\\
$^{137}$ Warsaw University of Technology, Warsaw, Poland\\
$^{138}$ Wayne State University, Detroit, Michigan, United States\\
$^{139}$ Yale University, New Haven, Connecticut, United States\\
$^{140}$ Yonsei University, Seoul, Republic of Korea\\
$^{141}$  Zentrum  f\"{u}r Technologie und Transfer (ZTT), Worms, Germany\\
$^{142}$ Affiliated with an institute covered by a cooperation agreement with CERN\\
$^{143}$ Affiliated with an international laboratory covered by a cooperation agreement with CERN.\\

\end{flushleft} 

\end{document}